\documentclass[preprint,12pt]{elsarticle}

\usepackage{amssymb}
\usepackage{amsmath}
\usepackage{graphicx}
\usepackage{here}
\usepackage{mathrsfs}
\usepackage{bbm}
\usepackage{physics}
\usepackage{bm}
\usepackage{comment}
\usepackage{mhchem}
\usepackage{siunitx}
\usepackage{url}
\usepackage{subcaption}
\usepackage{booktabs}

\journal{Computer Physics Communications}
\begin{document}

\begin{frontmatter}
  \title{First-principles analysis of in-plane anomalous Hall effect
  using symmetry-adapted Wannier Hamiltonians and multipole decomposition}
  \affiliation{organization={Department of Physics, Tohoku
    University},
    addressline={Aoba-ku},
    city={Sendai},
    postcode={980-8578},
  country={Japan}}
  \author{Hiroto Saito and Takashi Koretsune}

  \begin{abstract}
    The in-plane anomalous Hall effect occurs when magnetization lies
    within the same plane as the electric field and Hall current, and
    requires magnetic point groups lacking rotational or mirror symmetries.
    While it is observed in both Weyl semimetals and elemental
    ferromagnets, the microscopic role of higher-order multipoles
    remains unclear.
    Here, we develop a microscopic framework that combines
    time-reversal-symmetric Wannier functions with a symmetry-adapted
    multipole basis to decompose the first-principles Wannier Hamiltonian into
    electric, magnetic, magnetic toroidal, and electric toroidal multipoles.
    This approach allows us to rotate the magnetization rank by rank
    and quantify how each multipole affects the conductivity.
    Applying this framework to body-centered cubic iron, we find that
    high-rank magnetic and magnetic toroidal multipoles contribute
    with magnitudes comparable to magnetic dipoles, while magnetic
    toroidal 16-poles
    act with the opposite sign.
    Furthermore, based on this multipole analysis, we apply
    uniaxial strain along the [103] direction to control the dominant
    multipoles contributing to the 
    conductivity. The strain
    substantially modifies its angular dependence,
    demonstrating that
    multipole-resolved Hamiltonian
    engineering and magnetoelastic control serve as practical routes
    to predict and tune the in-plane anomalous Hall conductivity in
    simple ferromagnets.
  \end{abstract}
\end{frontmatter}

\section{Introduction}
The anomalous Hall effect (AHE) is a transverse transport phenomenon
in systems with broken time-reversal symmetry \cite{Nagaosa2010-tw,
Xiao2010-zr}.
Traditionally, AHE has
been understood as a response proportional to the magnetization,
where the Hall current flows perpendicular to the plane defined by
the magnetization and the electric field. However, in recent years,
Berry-curvature-based formulations and symmetry analyses have
significantly extended this conventional picture. In particular, many
theoretical and experimental studies have shown that AHE can appear
even in systems without a net magnetization, such as \ce{Mn3X} (\ce{X=Ir, Sn,
Ge, Pt, Rh}) \cite{Chen2014-co, Nakatsuji2015-lu, Nayak2016-cm,
  Ikhlas2017-mr, Zhang2017-bq, Yang2017-bq, Suzuki2017-dm, Liu2018-nz,
Zhao2019-oa, Chen2021-ua} and
altermagnets \cite{Naka2020-xi, Smejkal2022-fx, Reichlova2024-uc,
Takagi2024-gu}, where multipolar magnetic order or other
time-reversal-odd quantities play an essential role.

Along with these developments, the in-plane anomalous Hall effect
(IAHE) has attracted attention recently \cite{Liu2007-zs,
  Roman2009-fg, Tan2021-nd,
  Zhou2022-qh, Cao2023-yk, Kurumaji2023-qk,
  Nakamura2024-os, Miao2024-tc, Peng2024-oj, Nishihaya2025-tk, Wang2025-ys,
Liu2025-af}. IAHE
refers to an anomalous
Hall current that flows within the plane spanned by the magnetization
and the electric field. Compared with the conventional AHE geometry,
IAHE requires an additional magnetic point-group symmetry constraint
\cite{Tan2021-nd, Kurumaji2023-qk}. Therefore, its
observation can provide further information on low-symmetry magnetic
order and on topological band structures induced by spin-orbit coupling (SOC).

Initial experimental studies of IAHE focused on materials combining
strong SOC with topological band structures.
Hall responses consistent with IAHE have been reported in Weyl
semimetals and related systems such as \ce{EuCd_2Sb_2}
\cite{Nakamura2024-os}, \ce{Cd_3As_2}
\cite{Miao2024-tc}, \ce{SrRuO_3} \cite{Nishihaya2025-tk}, and
\ce{Co_3Sn_2S_2} \cite{Wang2025-ys}.
These observations were attributed to mechanisms such as effective
out-of-plane magnetization
components induced by orbital magnetization, or Weyl-point splitting
mediated by Berry curvature \cite{Nakamura2024-os}. More recently,
IAHE has also been
confirmed even in simple cubic ferromagnets such as \ce{Fe} and
\ce{Ni} \cite{Peng2024-oj}.

Particularly in cubic systems, a theoretical study predicted that
octupolar terms can cause the AHE vector to deviate from the
magnetization direction through higher-order spin-orbit perturbations
\cite{Liu2025-af}. Although this approach successfully explains the angular
dependence of IAHE, it relies on a phenomenological expansion with
respect to the spin magnetization, and does not explicitly connect
the microscopic terms related to higher-order multipoles or orbital
magnetism. Therefore, a microscopic treatment based on
first-principles-derived models is expected to provide complementary
and essential insights from the viewpoint of multipoles.

To quantify the IAHE as well as AHE in realistic material,
a tight-binding Hamiltonian from first-principles calculations is useful.
Maximally
localized Wannier functions (MLWFs) are widely used for this purpose,
as they enable accurate band interpolation and model construction
from density functional theory (DFT) by combining subspace disentanglement
\cite{Souza2001-hk} with spread minimization \cite{Marzari1997-rw}.
More recent extensions preserve key physical quantities
by, for example, optimizing Wannier functions with respect to their
projectability onto pseudo-atomic orbitals \cite{Qiao2023-mx},
constructing Wannier functions closest to atomic orbitals
\cite{Ozaki2024-ou, Oiwa2025-rm}, or employing symmetry-adapted
Wannier functions (SAWF) that enforce site-symmetry constraints
\cite{Sakuma2013-rq, Koretsune2023-ev}.
For ferromagnets, we previously developed SAWF and introduced the time-reversal
symmetric Wannier (TRS-Wannier) method, which
combines TR-symmetric gauge fixing with spin-group symmetry
constraints to capture the angular dependence of magnetic anisotropy
energy with high accuracy \cite{Saito2024-vw}.
In parallel, a symmetry-adapted multipole basis (SAMB) enables
decomposition of Hermitian operators into electric, magnetic,
magnetic toroidal, and electric toroidal types classified by rank and
the point-group irreducible representations \cite{Kusunose2023-fe,
Inda2024-oz, Hayami2024-ab, Oiwa2025-rm, Oiwa2025-td}.

Here, we combine the TRS-Wannier with SAMB analysis to
quantitatively evaluate the symmetry of Hamiltonians derived from
first-principles calculations.
This approach enables us to elucidate the microscopic origins of the
IAHE in ferromagnets.
In bcc \ce{Fe}, we analyze the angular dependence of the anomalous
Hall conductivity (AHC) under magnetization rotation by comparing
theoretical formulas,
DFT calculations, and multipole models.

\section{Method}

\subsection{In-plane anomalous Hall effect}
\label{sec:iahe}

We evaluate intrinsic AHC using linear response theory
\cite{Xiao2010-zr}. Current density $J_i$ is given by
\begin{align}
  & J_i = \sigma_{ij} E_j,
  \\
  & \sigma_{ij} = -\frac{e^2}{\hbar} \int_{\mathrm{BZ}} \frac{d^3 k}{8\pi^3}
  \, \Omega_{ij} (\bm{k}),   \\
  & \Omega_{ij} (\bm{k}) = -2 \hbar^2 \mathrm{Im} \sum_{n}^{\mathrm{occ.}}
  \sum_{m}^{\mathrm{unocc.}}
  \frac{\mel{\psi_{n\bm{k}}}{v_i}{\psi_{m\bm{k}}}
  \mel{\psi_{m\bm{k}}}{v_j}{\psi_{n\bm{k}}}}{(\varepsilon_{n\bm{k}} -
  \varepsilon_{m\bm{k}})^2},
\end{align}
where $i,j$ label Cartesian components, $\bm{E}$ is the applied electric field,
$\varepsilon_{n\bm{k}}$ and $\psi_{n\bm{k}}$ are the band energy
eigenvalues and Bloch eigenstates at wave vector $\bm{k}$, and
$v_i=(1/\hbar)\,\partial H/\partial k_i$ is the velocity operator.
Here, $\sigma_{ij}$ is an antisymmetric tensor
($\sigma_{ij}=-\sigma_{ji}$).

The anomalous Hall vector $\bm{\sigma}$ is defined as the pseudovector
$\sigma_k = \frac{1}{2} \epsilon_{ijk} \sigma_{ij}$, namely,
\begin{align}
  \bm{\sigma} & = (\sigma_{yz}, \sigma_{zx}, \sigma_{xy}).
\end{align}
If the magnetization direction $\bm{\hat{M}}$ is oriented along a
high-symmetry direction, $\bm{\sigma}$ is constrained to be parallel to it.
One can verify that $\bm{\sigma}$ transforms as a pseudovector.
That is, when we consider rotation matrices, $R_{ij} \in \mathrm{SO}(3)$,
$\sigma_{ij}$ transforms as
\begin{align}
  \sigma_{ij}' = R_{ip} R_{jq} \sigma_{pq},
\end{align}
and thus, we can obtain
\begin{align}
  \sigma_k' & = \frac{1}{2} \epsilon_{kij} R_{ip} R_{jq} \sigma_{pq}
  \nonumber  \\
  & = \frac{1}{2} \det(R) R_{kr} \epsilon_{rpq} \sigma_{pq} \nonumber \\
  & = \det(R) R_{kr} \sigma_r.
\end{align}
Here, we have used the following identity relation for rotation matrices:
\begin{align}
  \epsilon_{kij} R_{ip} R_{jq} = \det(R) R_{kr} \epsilon_{rpq}.
\end{align}

\begin{figure}[H]
  \centering
  \includegraphics[width=0.5\linewidth]{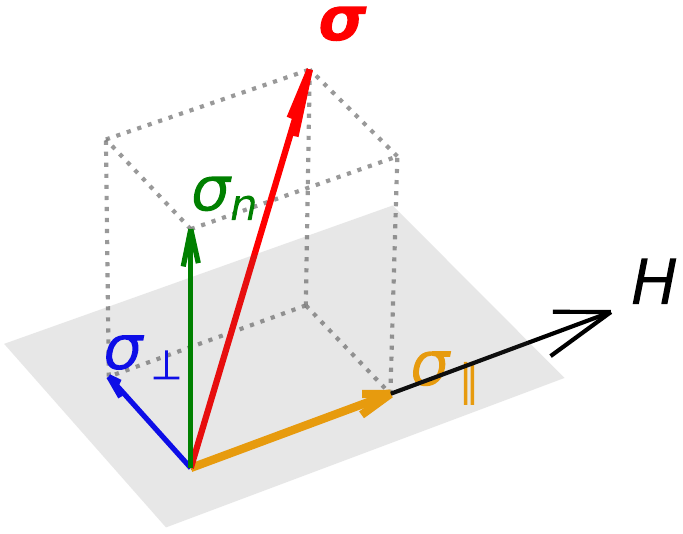}
  \caption{Decomposition of the anomalous Hall vector $\bm{\sigma}$
    into the components parallel to the magnetization
    ($\sigma_\parallel$), in-plane perpendicular ($\sigma_\perp$), and
    out-of-plane ($\sigma_{\bm n}$). The gray plane represents the
    plane formed by rotating $\bm{H}$.
  }
  \label{fig:sigma_vec}
\end{figure}
We decompose $\bm{\sigma}$ into three components with respect to the
magnetization direction $\bm{\hat{M}}$. When the magnetization
rotates within a plane with unit normal $\bm{n}$, we define the
in-plane unit vector orthogonal to $\bm{\hat{M}}$ as
$\bm{e}_\perp=\bm{n}\times\bm{\hat{M}}$. The components are then
\begin{align}
  \sigma_\parallel & = \bm{\sigma} \cdot \bm{\hat{M}}, \\
  \sigma_\perp     & = \bm{\sigma} \cdot \bm{e}_\perp, \\
  \sigma_{\bm{n}}  & = \bm{\sigma} \cdot \bm{n}.
\end{align}
Here, $\sigma_\parallel$ corresponds to the conventional AHC, while
$\sigma_\perp$ and $\sigma_{\bm{n}}$
represent the IAHE components (Fig.~\ref{fig:sigma_vec}).
In the experiment, the measurement direction of $\bm{\sigma}$ is fixed while
the magnetization is rotated. As a result, $\sigma_\parallel$ and $\sigma_\perp$
rotate with the magnetization, whereas the direction of $\sigma_{\bm{n}}$
remains invariant.
For example, when the magnetization rotates by angle $\psi$ in the $xy$ plane,
the components transform as
\begin{align}
  \begin{pmatrix}
    \sigma_\parallel \\
    \sigma_\perp     \\
    \sigma_{\bm{n}}
  \end{pmatrix}
  =
  \begin{pmatrix}
    \cos\psi  & \sin\psi & 0 \\
    -\sin\psi & \cos\psi & 0 \\
    0         & 0        & 1
  \end{pmatrix}
  \begin{pmatrix}
    \sigma_{yz} \\
    \sigma_{zx} \\
    \sigma_{xy}
  \end{pmatrix}
  =
  \begin{pmatrix}
    \sigma_{\bm{e}\perp, z}  \\
    \sigma_{z, \bm{\hat{M}}} \\
    \sigma_{xy}
  \end{pmatrix}.
\end{align}
Therefore, it is possible to measure the
magnetization-rotation dependence only of $\sigma_{\bm{n}}$.

\begin{figure}[H]
  \begin{subfigure}[b]{0.48\linewidth}
    \centering
    \includegraphics[width=\linewidth]{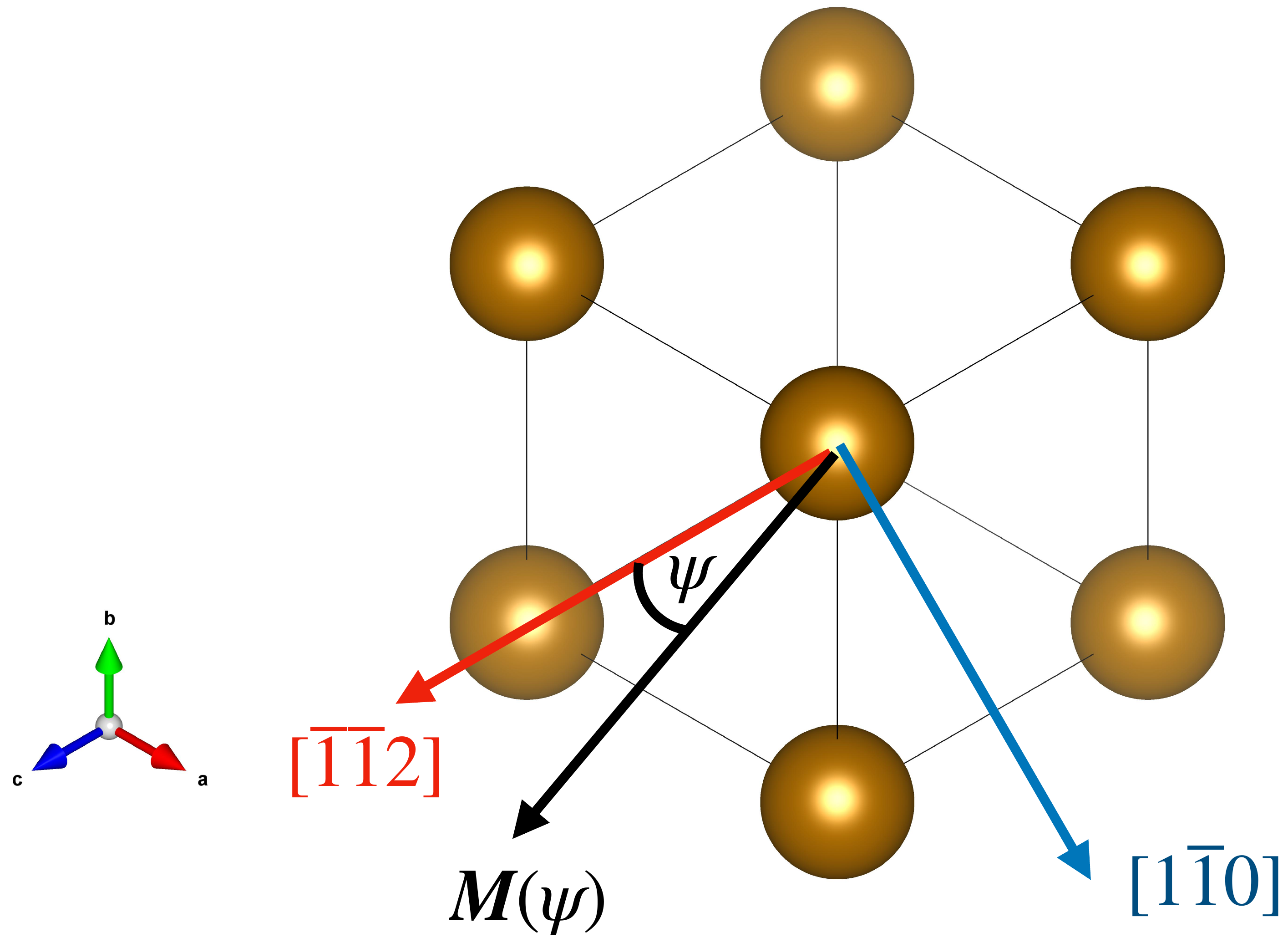}
    \caption{Magnetization rotation within the $(111)$ plane from
    [$\overline{1}\overline{1}2$] to [$1\overline{1}0$].}
  \end{subfigure}\hfill
  \begin{subfigure}[b]{0.48\linewidth}
    \centering
    \includegraphics[width=\linewidth]{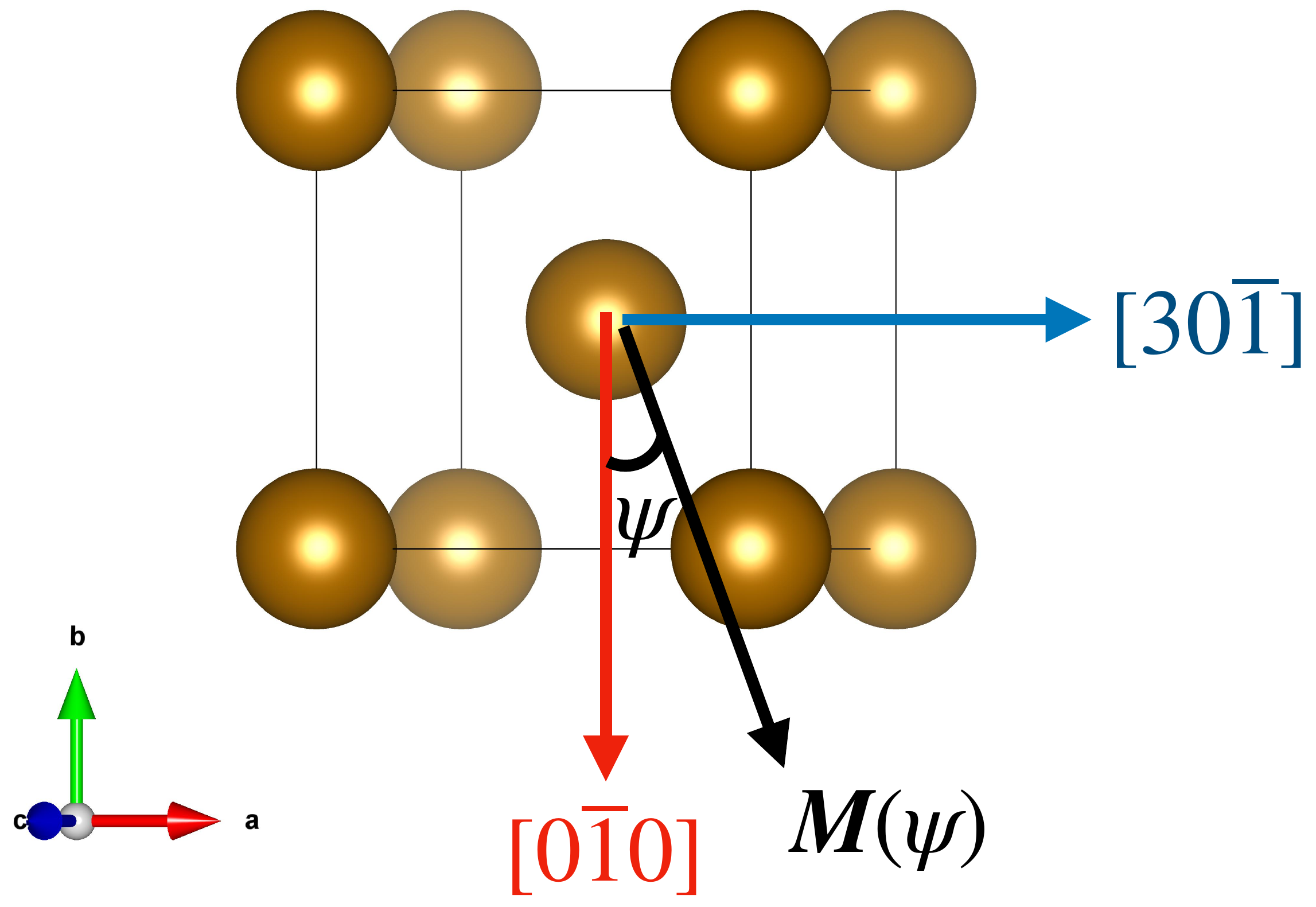}
    \caption{Magnetization rotation within the $(103)$ plane from
    [$0\overline{1}0$] to [$30\overline{1}$].}
  \end{subfigure}
  \caption{Schematic definitions of the magnetization rotation planes
  and the azimuthal angle $\psi$.}
  \label{fig:bcc_planes}
\end{figure}

The magnetization-direction dependence of $\bm{\sigma}$ is
phenomenologically expanded in terms of multipoles in spin space
\cite{Liu2025-af}:
\begin{align}
  \bm{\sigma} &= p_{ij} \hat{M}_j \bm{e}_i + o_{ijkl} \hat{M}_j
  \hat{M}_k \hat{M}_l \bm{e}_i + \cdots, \\
  p_{ij} &= \frac{3}{4\pi} \int \sin\theta d\phi \sigma_i(\theta,
  \phi) \hat{M}_j \\
  o_{ijkl} &= \frac{7}{8\pi} \int \sin\theta d\phi \sigma_i(\theta, \phi)
  [5\hat{M}_j \hat{M}_k \hat{M}_l - \hat{M}_j\delta_{kl} -
  \hat{M}_k\delta_{jl} - \hat{M}_l\delta_{jk}],
\end{align}
where $\bm{e}_i$ is the unit vector along the $i$-th Cartesian axis.
For cubic-symmetric systems, this reduces to
\begin{align}
  \bm{\sigma}
  = \alpha
  \begin{pmatrix}
    M_x \\
    M_y \\
    M_z
  \end{pmatrix}
  + \beta
  \begin{pmatrix}
    M_x^3 \\
    M_y^3 \\
    M_z^3
  \end{pmatrix}, \label{eq:sigma_cubic}
\end{align}
with $\alpha = p_{xx} - \frac{3}{2}o_{xxxx}$ and $\beta =\frac{5}{2}o_{xxxx}$.
Based on this expression, we derive the angular dependence of
$\sigma_\parallel$, $\sigma_\perp$, and $\sigma_{\bm{n}}$ for
magnetization rotation in the $(111)$ and $(103)$ planes
(Fig.~\ref{fig:bcc_planes}).

For a rotation by angle $\psi$, angle-dependence of each component is
given as follows:
\begin{itemize}
  \item $(111)$ plane rotation from $[\overline{1}\overline{1}2] \to
    [1\overline{1}0]$
    \begin{align}
      \sigma_\parallel & = \alpha + \frac{\beta}{2},
      \label{eq:sigma_parallel_111}\\
      \sigma_\perp     & = 0,
      \label{eq:sigma_perp_111}\\
      \sigma_{\bm{n}}  & = \frac{\sqrt{6}}{6}\beta\cos3\psi.
      \label{eq:sigma_n_111}
    \end{align}
  \item $(103)$ plane rotation from $[0\overline{1}0] \to [30\overline{1}]$
    \begin{align}
      \sigma_\parallel & = \alpha + \beta\left(\frac{273}{400} +
      \frac{9}{100}\cos2\psi + \frac{91}{400}\cos4\psi\right),
      \label{eq:sigma_parallel_103}\\
      \sigma_\perp     & = -\frac{\beta}{400}\left(18\sin2\psi +
      91\sin4\psi\right),
      \label{eq:sigma_perp_103}\\
      \sigma_{\bm{n}}  & = \frac{6\beta}{25}\sin^3\psi.
      \label{eq:sigma_n_103}
    \end{align}
\end{itemize}

\subsection{Symmetry-adapted multipole basis}
\label{sec:samb}

We construct SAMB to represent all electronic degrees of freedom with
a complete and orthonormal basis.
SAMB is defined as the direct product of atomic multipole
$\mathbb{X}^{(\mathrm{a})}$ that describes atomic orbital parts and
site/bond-cluster multipole $\mathbb{Y}^{(\mathrm{s/b})}$ that
corresponds to sublattices or bonding clusters \cite{Kusunose2020-oa,
Kusunose2023-fe}.
Specifically,
\begin{align}
  &\mathbb{Z}(l, \Gamma, n, \gamma | s, k) \nonumber\\
  & = \sum_{\substack{\Gamma_1, n_1, \gamma_1 \\
  \Gamma_2, n_2, \gamma_2}}
  C(X,Y |Z)
  ~\mathbb{X}^{(\mathrm{a})}(l_1, \Gamma_1, n_1, \gamma_1 | s, k)
  \otimes \mathbb{Y}^{(\mathrm{s/b})}(l_2, \Gamma_2, n_2, \gamma_2),
\end{align}
where $l$ is rank, $\Gamma$ is the space group irreducible
representation, $n$ is multiplicity,
$\gamma$ is the component index within irreducible representation
(e.g., $u,v$ for $\mathrm{E}$ representation, $x,y,z$ for
$\mathrm{T}$ representation),
$s$ indicates spin degrees of freedom ($s=0$ or $1$), $k$ represents
spin components ($k=0$ for $s=0$ and $k=-1,0,1$ for $s=1$),
and $C(X,Y | Z)$ is the Clebsch-Gordan coefficient.
These indices are omitted when
redundant, and we sometimes collectively denote the set of indices by $i$.

Each multipole is classified as one of four types according to its
transformation properties under time-reversal and spatial-inversion
symmetries: electric ($\mathbb{Q}$), magnetic ($\mathbb{M}$),
magnetic toroidal ($\mathbb{T}$), and electric toroidal ($\mathbb{G}$).
SAMB is orthonormal and complete, satisfying
\begin{align}
  & \braket{\mathbb{Z}_i}{\mathbb{Z}_j}
  =\mathrm{tr}[\mathbb{Z}_i^\dagger \mathbb{Z}_j] = \delta_{ij}, \\
  & \left[ \sum_i \ketbra{\mathbb{Z}_i}{\mathbb{Z}_i} \right]_{aa', bb'}
  = \sum_i [\mathbb{Z}_i]_{aa'} [\mathbb{Z}_i^*]_{bb'}
  = \delta_{ab} \delta_{a'b'}.
\end{align}
This enables unique decomposition of the Hamiltonian $H$, with
multipole decomposition defined as
\begin{align}
  & H^{\mathrm{MD}} = \sum_{i} z_i \mathbb{Z}_i, \\
  & z_i = \mathrm{tr} [\mathbb{Z}^\dagger_i H] ,
\end{align}
where coefficient $z_i$ represents contributions from each multipole component.

For quantitative evaluation of symmetry preservation of Wannier
Hamiltonians, we define energy eigenvalue differences before and
after decomposition as
\begin{align}
  & \Delta_{\mathrm{energy}}
  = \frac{1}{N_{\bm{k}} N_{\mathrm{W}}}
  \sum_{n\bm{k}}
  |\varepsilon_{n\bm{k}}
  - \varepsilon^{\mathrm{MD}}_{n\bm{k}}| ,
  \label{eq:delta_energy}
\end{align}
where $N_{\bm{k}}$ denotes the number of $\bm{k}$-points sampled in
the Brillouin zone, $N_{\mathrm{W}}$ is the number of Wannier bands,
and $\varepsilon^{\mathrm{MD}}_{n\bm{k}}$ is the energy eigenvalue of
$H^{\mathrm{MD}}$.

\subsection{Time-reversal symmetric Wannier method and multipole rotation}
\label{sec:trs-multipole}
We use the TRS-Wannier method for a high-precision analysis of
the AHE during magnetization rotation.
Expanding the Hamiltonian into a multipole basis and introducing
rank-wise rotation operations enables a quantitative evaluation of the
contribution of each multipole component to the AHC.
This section describes specific formulation and definitions of
rotation operations.

In the TRS-Wannier method, Wannier function basis preserves
time-reversal symmetry during Wannierization \cite{Saito2024-vw}.
Choosing a spin-quantization axis perpendicular to the magnetization
(e.g., take $x\perp z$) makes the $S_x=\pm\tfrac12$ Wannier functions
identical in real space, because site symmetries relate the two spin
states. For example, in cubic crystals a $C_{2z}$ rotation acts similarly to
time reversal for an in-plane axis. As a result, the two spins share
the same centers and spreads and form time-reversal-symmetric pairs.

We also apply spin group symmetry constraints to the Hamiltonian
\cite{Liu2022-hy, Chen2025-py, Liu2025-af}.
The ferromagnetic spin group is the direct product of spin-only group
$\infty 2'$ and crystal point group, justified in the negligible SOC limit.
Decomposing the Hamiltonian into time-reversal symmetric and
antisymmetric parts yields:
\begin{align}
  H^{\mathrm{MD}}
  = & H^{\mathrm{s}} + H^{\mathrm{a}} \nonumber                            \\
  = & \left( \sum_{i} q_i \mathbb{Q}_i + \sum_{i} g_i \mathbb{G}_i \right)
  + \left( \sum_{i} m_i \mathbb{M}_i + \sum_{i} t_i \mathbb{T}_i
  \right). \label{eq:ham_decomp}
\end{align}
Since SOC is time-reversal symmetric, it belongs to the first term
$H^{\mathrm{s}}$. The second term $H^{\mathrm{a}}$ represents the
Hamiltonian without SOC and should satisfy spin-only group symmetry.
From the Hermiticity of the Hamiltonian, $H^{\mathrm{a}}$ can
generally be written as
\begin{align}
  H^{\mathrm{a}} = i A_0 + \bm{A} \cdot \bm{s}, \label{eq:ham_asym}
\end{align}
where $\bm{s} = (s_x, s_y, s_z)$ is the Pauli matrix vector.

The spin-only group $\infty 2'$ consists of spin rotations $C^s_{\psi
z}$ around the $z$-axis (magnetization direction) with arbitrary
angle $\psi$ $(0 \leq \psi < 2\pi)$, composite operations $T
C^s_{2x}$ combining spin rotation about $x$-axis, $C^s_{2x}$, with
time reversal operation, $T$, and their products, forming a continuous group.
The constraint imposed by $C^s_{\psi z}$ symmetry gives
$A_x = A_y = 0$,
while the constraint by $T C^s_{2x}$ yields
$A_0 = 0$.
Therefore, $H^{\mathrm{a}}$ is limited to
\begin{align}
  H^{\mathrm{a}} = A_z s_z,
\end{align}
i.e., only the magnetization direction component.

Next, let us consider the spin rotation for the TRS Hamiltonian.
The TRS Hamiltonian that includes only the $s_z$ component in
$H^{\mathrm{a}}$ is expressed as,
\begin{align}
  H^{\mathrm{TRS}}
  = & \left( \sum_{i} q_i \mathbb{Q}_i + \sum_{i} g_i \mathbb{G}_i \right)
  +\mathrm{tr}\left[s_z^\dagger \left( \sum_{i} m_i \mathbb{M}_i +
  \sum_{i} t_i \mathbb{T}_i \right) \right] s_z, \label{eq:ham_trs}
\end{align}
where the trace is taken over spin space.
Using the spinor rotation matrix
\begin{align}
  U^{\mathrm{s}}(\theta, \phi) =
  \begin{pmatrix}
    \cos \theta/2            & e^{-i\phi} \sin \theta/2 \\
    -e^{i\phi} \sin \theta/2 & -\cos \theta/2
  \end{pmatrix},
\end{align}
we can rotate magnetization from $z$-axis to general direction
$(\theta, \phi)$, and  obtain $H^{\mathrm{TRS}}(\theta, \phi)$ as
\begin{align}
  H^{\mathrm{TRS}} (\theta, \phi)
  = & \left( \sum_{i} q_i \mathbb{Q}_i + \sum_{i} g_i \mathbb{G}_i
  \right) \nonumber\\
  + & \mathrm{tr}\left[s_z^\dagger \left( \sum_{i} m_i \mathbb{M}_i +
  \sum_{i} t_i \mathbb{T}_i \right) \right]
  \begin{pmatrix}
    \cos\theta           & e^{-i\phi} \sin\theta \\
    e^{i\phi} \sin\theta & -\cos\theta
  \end{pmatrix}.
\end{align}
For the rank-\(l\) multipole rotation, we define the Hamiltonian with
only the rank-\(l\) multipole components rotated,
\(H^{\mathrm{TRS}}_{l}(\theta,\phi)\), as
\begin{align}
  H^{\mathrm{TRS}}_{l}(\theta, \phi)
  = & \left( \sum_{i} q_i \mathbb{Q}_i + \sum_{i} g_i \mathbb{G}_i
  \right) \nonumber \\
  + & \mathrm{tr}\left[s_z^\dagger \left( \sum_{i \in \mathrm{rank}~l}
      m_i \mathbb{M}_i
  + \sum_{i \in \mathrm{rank}~l} t_i \mathbb{T}_i \right) \right]
  \begin{pmatrix}
    \cos\theta           & e^{-i\phi} \sin\theta \\
    e^{i\phi} \sin\theta & -\cos\theta
  \end{pmatrix}.
\end{align}

\subsection{Calculation details}
\label{sec:calc-details}
\begin{figure}[H]
  \centering
  \includegraphics[width=0.4\linewidth]{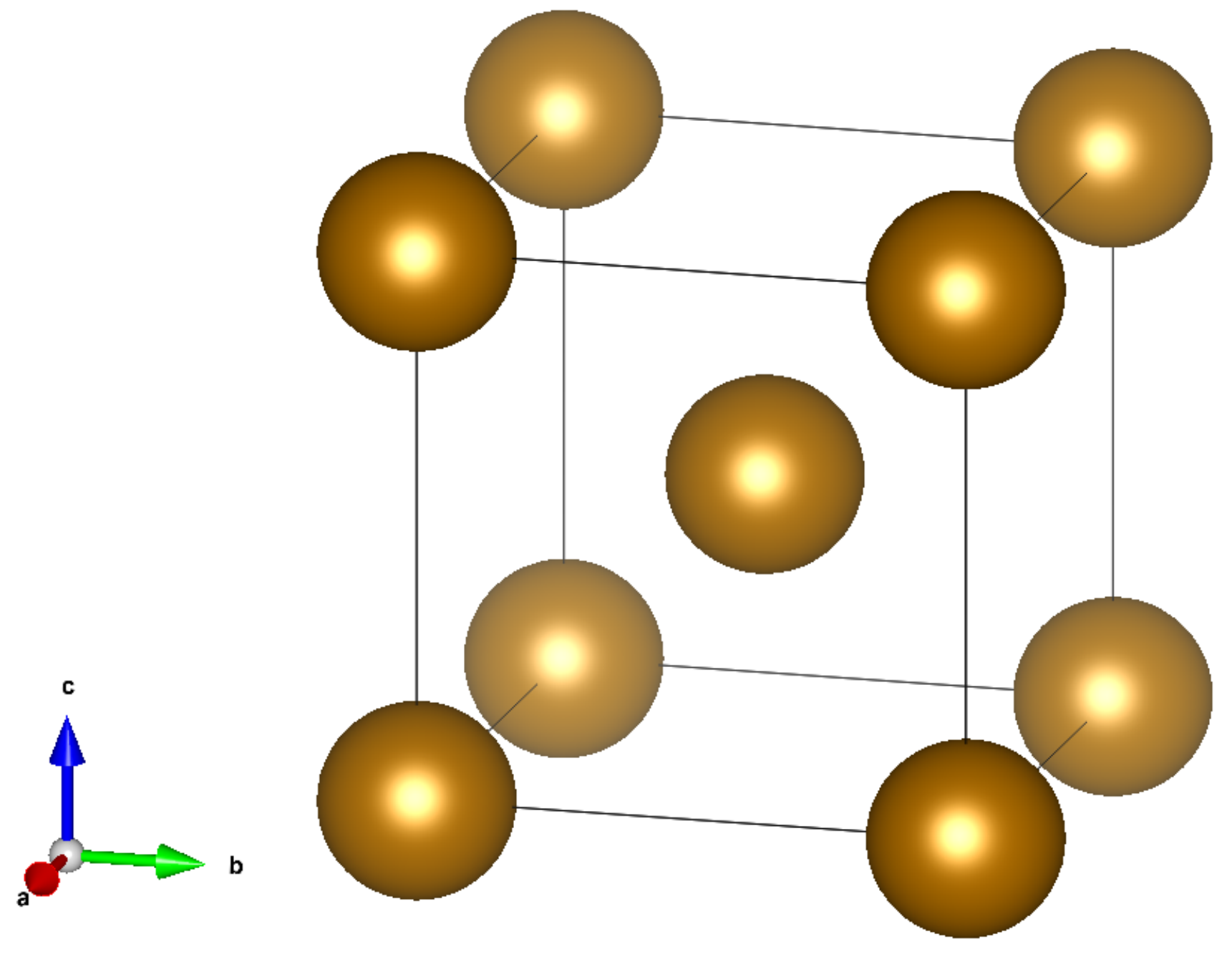}
  \caption{Crystal structures of body-centered cubic iron (bcc $\ce{Fe}$).}
  \label{fig:structure}
\end{figure}
We performed first-principles calculations using the {\sc{Quantum
ESPRESSO}} package \cite{Giannozzi2009-zd,Giannozzi2017-yu} with a
plane-wave basis set and pseudopotential methods.
The resulting electronic structures were used to construct
tight-binding models via the {\sc SymWannier} program \cite{Koretsune2023-ev}.
For bcc $\ce{Fe}$ (space group No.~229), the lattice constant was set
to $a = 2.868~\si{\angstrom}$ (Fig.~\ref{fig:structure}).
The plane-wave energy cutoffs for the wave functions and the charge
density were $64.0~\mathrm{Ry}$ and $782.0~\mathrm{Ry}$, respectively.
Exchange-correlation effects were treated within the
Perdew-Burke-Ernzerhof (PBE) generalized gradient approximation (GGA)
\cite{Perdew1996-tj}.
Ultrasoft pseudopotentials from {\sc PSLibrary}
\cite{Dal_Corso2014-hd} were employed.
SOC was included in the non-self-consistent field calculations,
and the Fermi energy used in all subsequent analyses was taken from this step.

Wannierization was carried out in a one-shot procedure, projecting
onto localized $s$, $p$, and $d$ orbitals at atomic sites, resulting
in 36 spinor Wannier functions per atom. An $8 \times 8 \times 8$
$k$-mesh was used, corresponding to 59 inequivalent $k$ points under symmetry.
The inner energy window for disentanglement was chosen to include all
$3d$ states. Importantly, the spin quantization axis during
Wannierization was set perpendicular to the magnetization direction
obtained from first-principles calculations. This ensured differences
in Wannier centers of less than \(10^{-6}\si{\angstrom}\) and
differences in spreads of less than \(10^{-8}\si{\angstrom^2}\) for
all up/down spinor pairs.

SAMB operators were generated with the {\sc MultiPie} program
\cite{Kusunose2023-fe}. We constructed operators for all irreducible
representations and included bond terms through the 35th-nearest
neighbor, thereby covering every bond present in the Wannier
Hamiltonian. Retaining the full set of irreducible representations
$({\mathrm{A_{1g}}}\oplus{\mathrm{A_{2g}}}\oplus{\mathrm{E_{g}}}
  \oplus{\mathrm{T_{1g}}}\oplus{\mathrm{T_{2g}}}\oplus{\mathrm{A_{1u}}}
  \oplus{\mathrm{A_{2u}}}\oplus{\mathrm{E_{u}}}
\oplus{\mathrm{T_{1u}}}\oplus{\mathrm{T_{2u}}})$
yields a total of 345,708 SAMB matrices.

AHC calculations were conducted with the {\sc WannierBerri} code
\cite{Tsirkin2021-xq}. Starting from a uniform $10^6$ $k$-mesh, we
performed 20 adaptive interpolation loops with a refinement factor of
50. The final calculations used approximately $N_k \approx 1.32
\times 10^8$ $k$ points, confirming convergence.

\section{Result}
\subsection{Multipole decomposition of Hamiltonians and
analysis of dominant components}
\subsubsection{Evaluation of Hamiltonians using
Symmetry-adapted multipole basis}
\begin{figure}[H]
  \centering
  \includegraphics[width=\linewidth]{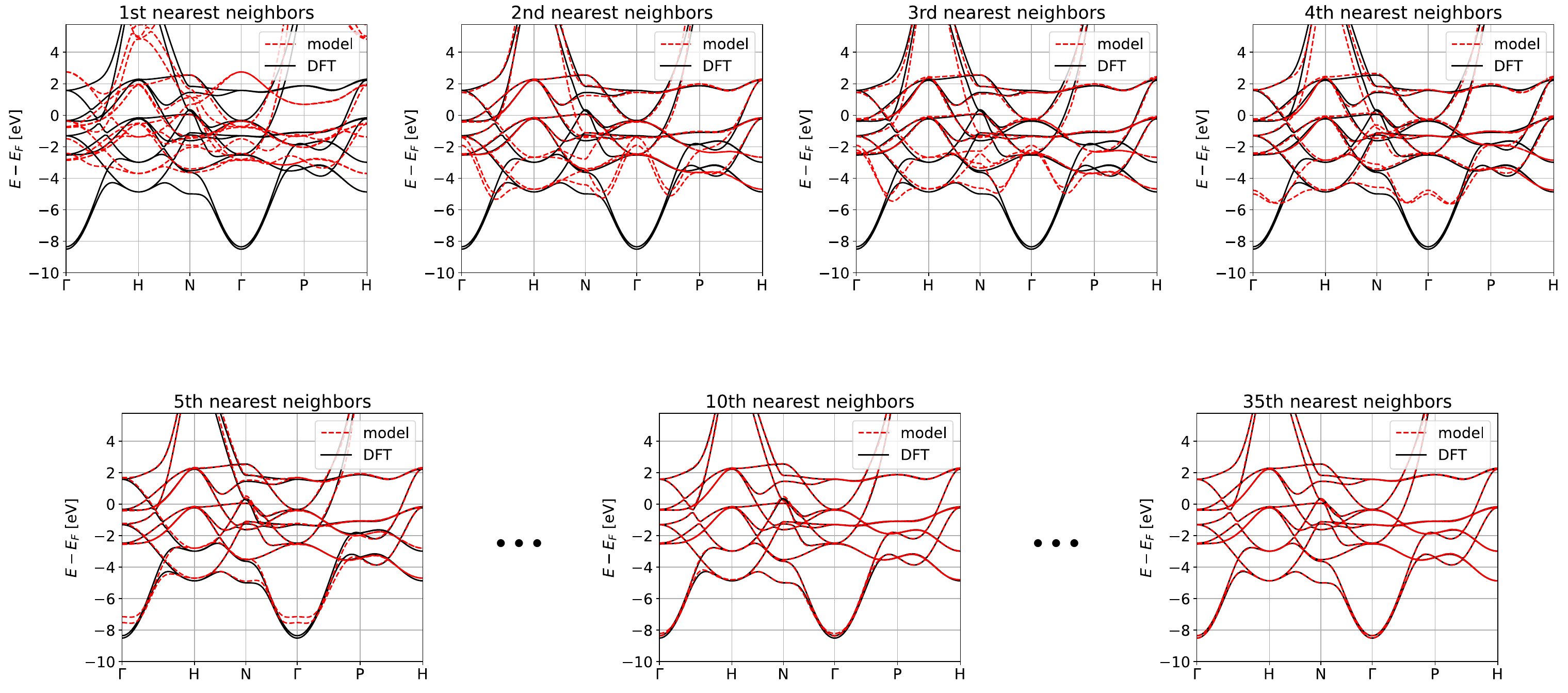}
  \caption{Dependence of the band structure on the bond range.
    Considering neighbors up to the 35th-nearest neighbors leads to
  convergence with the DFT results.}\label{fig:band_bond}
\end{figure}

We verify the completeness of the multipole decomposition of the
TRS-Wannier Hamiltonian using the SAMB framework.
Including multipoles up to the 35th-nearest neighbors ensures that the
band structure converges to the DFT results (Fig.~\ref{fig:band_bond}).

Beyond completeness, it is important to verify which irreducible
representations actually appear in the Hamiltonian, as this
substantiates the physical soundness of the decomposition.
In this case, angular momentum operators $\bm{L}$ and $\bm{S}$
belong to $\mathrm{T_{1g}}$ representation.
The bilinear, $\bm{L} \cdot \bm{S}$, decomposes as
$\mathrm{T_{1g}}\otimes\mathrm{T_{1g}}
=\mathrm{A_{1g}}\oplus\mathrm{E_g}\oplus\mathrm{T_{2g}}$.
Therefore, besides $\mathrm{A_{1g}}$ and $\mathrm{T_{1g}}$,
the Hamiltonian can contain
$\mathrm{E_g}$ and $\mathrm{T_{2g}}$ components as well.
Table~\ref{tab:delta_energy} summarizes the energy difference between
the original and multipole decomposed Hamiltonians,
$\Delta_{\mathrm{energy}}$, defined in Eq.~(\ref{eq:delta_energy}).
Here, we consider Wannier Hamiltonians obtained by the original Wannier90
and the TRS-Wannier methods introduced in Sec.\ \ref{sec:trs-multipole}.
To check this convergence reliably, we modified both the original
Wannier90 and the TRS-Wannier codes so that the output precision was
increased to 16 digits after the decimal point.
The first row corresponds to the multipole decomposition only using
$\mathrm{A_{1g}} \oplus \mathrm{T_{1g}}$, the
minimal set containing the identity and primary magnetic components.
The second row adds the SOC-induced $\mathrm{E_g}$ and
$\mathrm{T_{2g}}$ representations generated by $\mathrm{T_{1g}} \otimes
\mathrm{T_{1g}}$.
Including $\mathrm{E_g}$ and $\mathrm{T_{2g}}$ reduces
$\Delta_{\mathrm{energy}}$ to the order of $10^{-12}$ eV for TRS-Wannier,
yielding an almost fully reconstructed Hamiltonian.
TRS-Wannier attains about three orders of magnitude better
symmetry-preserving accuracy than ordinary Wannier90 for the
truncated physically motivated set,
confirming the effectiveness of the symmetry constraints.
SAMB thus provides a quantitative metric for Hamiltonian symmetry
fidelity and a physically interpretable pathway for systematic improvement.

\begin{table}[H]
  \centering
  \caption{Comparison of multipole decomposition reproduction
  accuracy between ordinary Wannier90 method and TRS-Wannier method.}
  \label{tab:delta_energy}
  \begin{tabular}{lcc}
    \toprule
    & \multicolumn{2}{c}{$\Delta_{\mathrm{energy}}$ [eV]}
    \\
    \cmidrule(lr){2-3}
    Irreducible representation
    & Wannier90                                           &
    TRS-Wannier             \\
    \midrule
    $\mathrm{A_{1g}} \oplus \mathrm{T_{1g}}$
    & $1.595 \times 10^{-5}$                              & $1.607
    \times 10^{-5}$  \\
    $\mathrm{A_{1g}} \oplus \mathrm{T_{1g}} \oplus \mathrm{E_g}
    \oplus \mathrm{T_{2g}}$ & $1.415 \times 10^{-9}$
    & $2.024 \times 10^{-12}$ \\
    \bottomrule
  \end{tabular}
\end{table}

\subsubsection{Multipole decomposition
coefficients and physical interpretation}
\begin{figure}[H]
  \centering
  \includegraphics[width=0.7\linewidth]{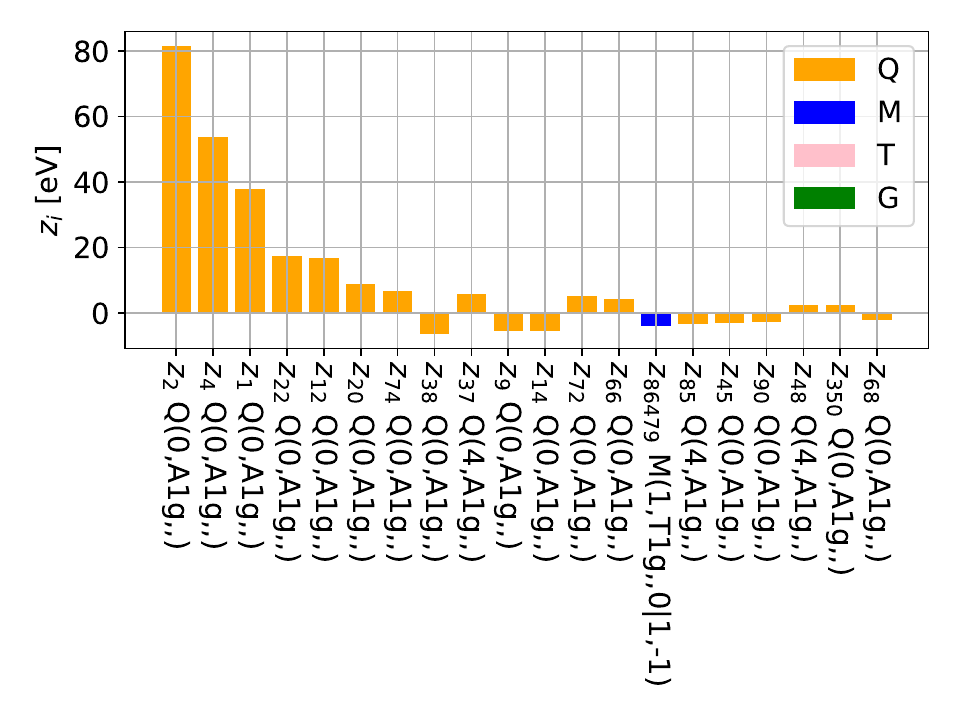}
  \caption{Top 20 coefficients in multipole decomposition of
  TRS-Wannier Hamiltonian.}\label{fig:zcoeff_all}
\end{figure}

We extract the highest-magnitude coefficients from the multipole decomposition
of the TRS-Wannier Hamiltonian. Figure \ref{fig:zcoeff_all} shows the top twenty
coefficients, indicating that the dominant contributions are mostly
electric multipoles $\mathbb{Q}$ belonging to the identity representation
$\mathrm{A_{1g}}$. Among the leading terms, $z_{86479}$
($\mathbb{M}(1,\mathrm{T_{1g}},,0|1,-1)$) is a $\mathrm{T_{1g}}$
magnetic dipole that constitutes the primary magnetic-ordering
component of the Fe $3d$ orbitals.

\begin{figure}[H]
  \centering
  \includegraphics[width=0.7\linewidth]{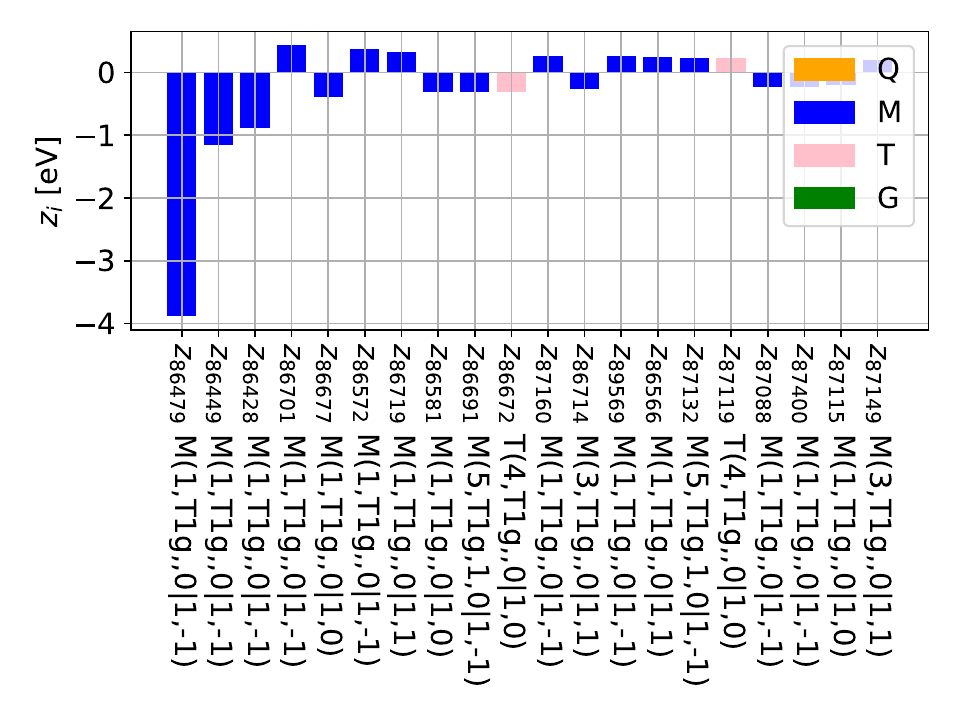}
  \caption{Top 20 coefficients in multipole decomposition of
    TRS-Wannier Hamiltonian (excluding electric multipoles
  $\mathrm{Q}$).}\label{fig:zcoeff_wo_q}
\end{figure}

\begin{figure}[H]
  \centering
  \includegraphics[width=0.8\linewidth]{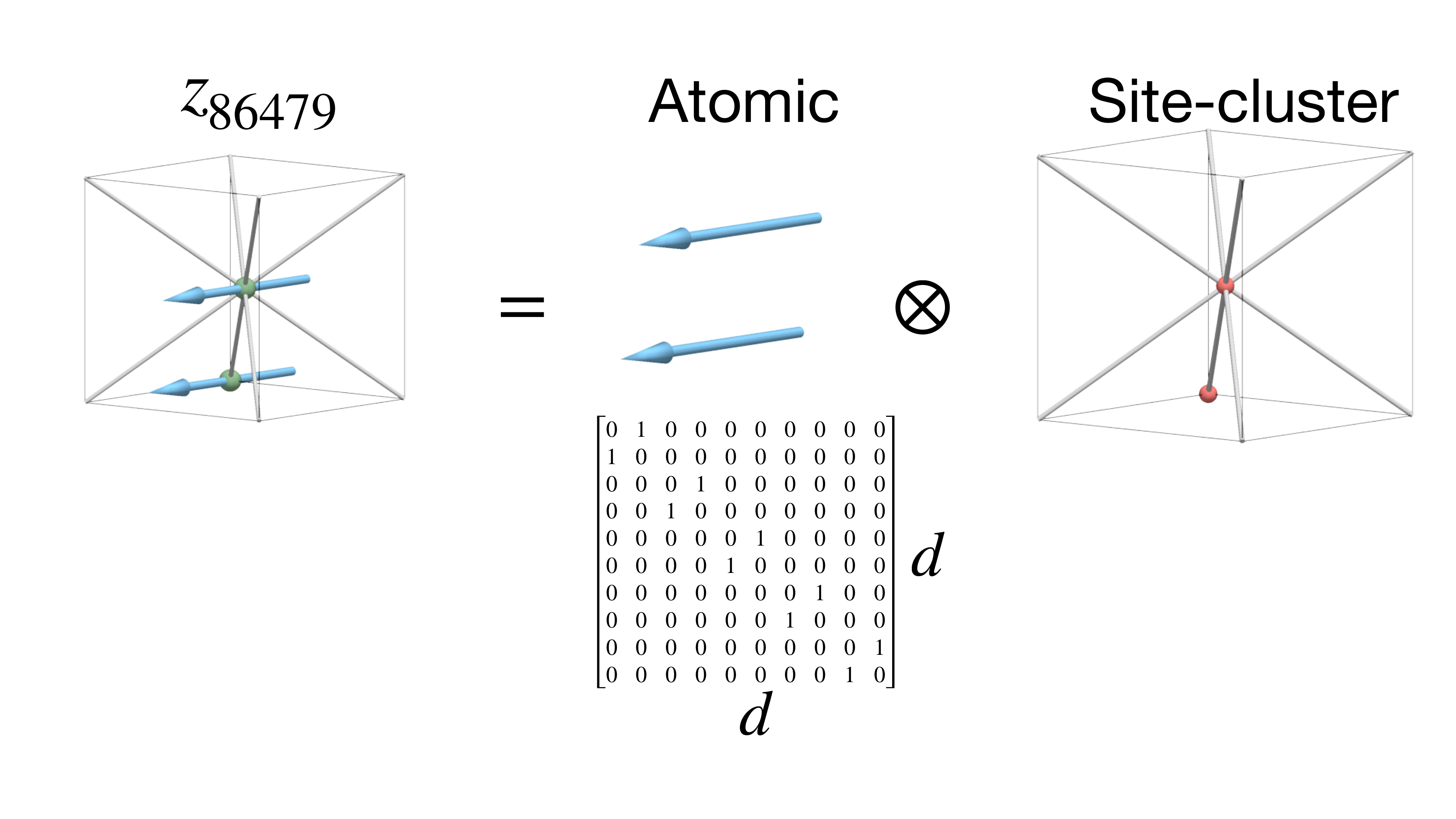}
  \caption{Schematic of SAMB $z_{86479}$
    ($\mathbb{M}(1,\mathrm{T_{1g}},,0|1,-1)$).
    The atomic magnetic dipole represents the magnetization-direction
    components of the $d$ orbitals, whereas the site-cluster electric
    monopole represents an intra-atomic contribution.
    All figures of schematic multipoles were prepared with {\sc
    QtDraw} \cite{Kusunose2023-fe}.
  \label{fig:MP86479}}
\end{figure}

Figure~\ref{fig:zcoeff_wo_q} shows the twenty largest coefficients in
the multipole decomposition with electric multipoles excluded,
thereby highlighting the leading magnetic order.
It is reasonable that they all belong to the same $\mathrm{T_{1g}}$
representation as magnetization.
The three largest non-electric contributions—$z_{86479}$, $z_{86449}$, and
$z_{86428}$—share the same factorization:
\begin{align}
  \mathbb{M}(1,\mathrm{T_{1g}},,0|1,-1)
  = \mathbb{M}^{(\mathrm{a})}(1,\mathrm{T_{1g}},,0|1,-1) \otimes
  \mathbb{Q}^{(\mathrm{s})}(0,\mathrm{A_{1g}},,). \label{eq:MP86479}
\end{align}
Here, the atomic multipole $\mathbb{M}^{(\mathrm{a})}$ represents the
components of magnetic dipoles arising from the $d$, $p$, and $s$
orbitals, whereas the site-cluster electric monopole
$\mathbb{Q}^{(\mathrm{s})}$ represents an
intra-atomic contribution (Fig.~\ref{fig:MP86479}).
In other words, these correspond to exchange splitting term,
$J_z\sigma_z$, for $d$, $p$, and $s$ orbitals.

\begin{figure}[H]
  \centering
  \includegraphics[width=0.8\linewidth]{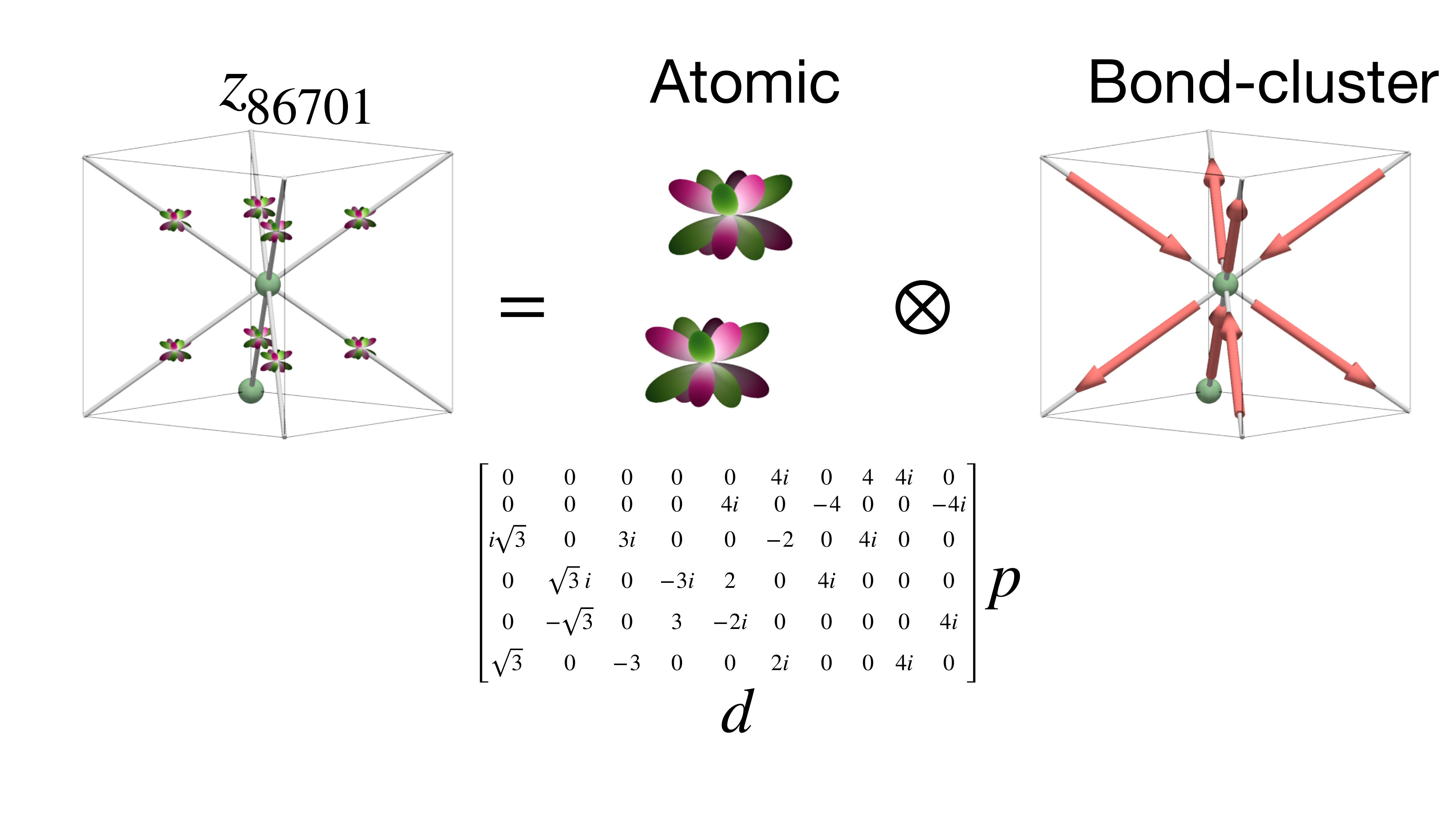}
  \caption{Schematic view of SAMB $z_{86701}$
    ($\mathbb{M}(1,\mathrm{T_{1g}},,0|1,-1)$).
    The atomic electric-toroidal 16-pole $\mathbb{G}$ represents
    $p$-$d$ order,
    while the bond-cluster magnetic-toroidal octupole $\mathbb{T}$ corresponds
  to purely imaginary nearest-neighbor hopping.}
  \label{fig:MP86701}
\end{figure}

By contrast, the coefficient of the fourth-largest magnetic-dipole in
Fig.~\ref{fig:zcoeff_wo_q},
$z_{86701}$ ($\mathbb{M}(1,\mathrm{T_{1g}},,0|1,-1)$), has a different
factorization:
\begin{align}
  \mathbb{M}(1,\mathrm{T_{1g}},,0|1,-1) =
  & \mathbb{G}^{(\mathrm{a})}(4,\mathrm{T_{2u}},,0|1,-1) \otimes
  \mathbb{T}^{(\mathrm{b})}(3,\mathrm{A_{2u}},,). \label{eq:MP86701}
\end{align}
Here, the atomic multipole corresponds to electric-toroidal 16-pole
$\mathbb{G}$ involving the $p$ and $d$ orbitals, while the
bond-cluster multipole describes magnetic-toroidal
octupole hopping $\mathbb{T}$ that is purely imaginary on nearest-neighbor bonds
(Fig.~\ref{fig:MP86701}).
Thus, even in bcc $\ce{Fe}$, nearest-neighbor $p$-$d$ hopping
accompanied by simultaneous spin- and
orbital-changes is appreciable.

\begin{figure}[H]
  \centering
  \includegraphics[width=0.8\linewidth]{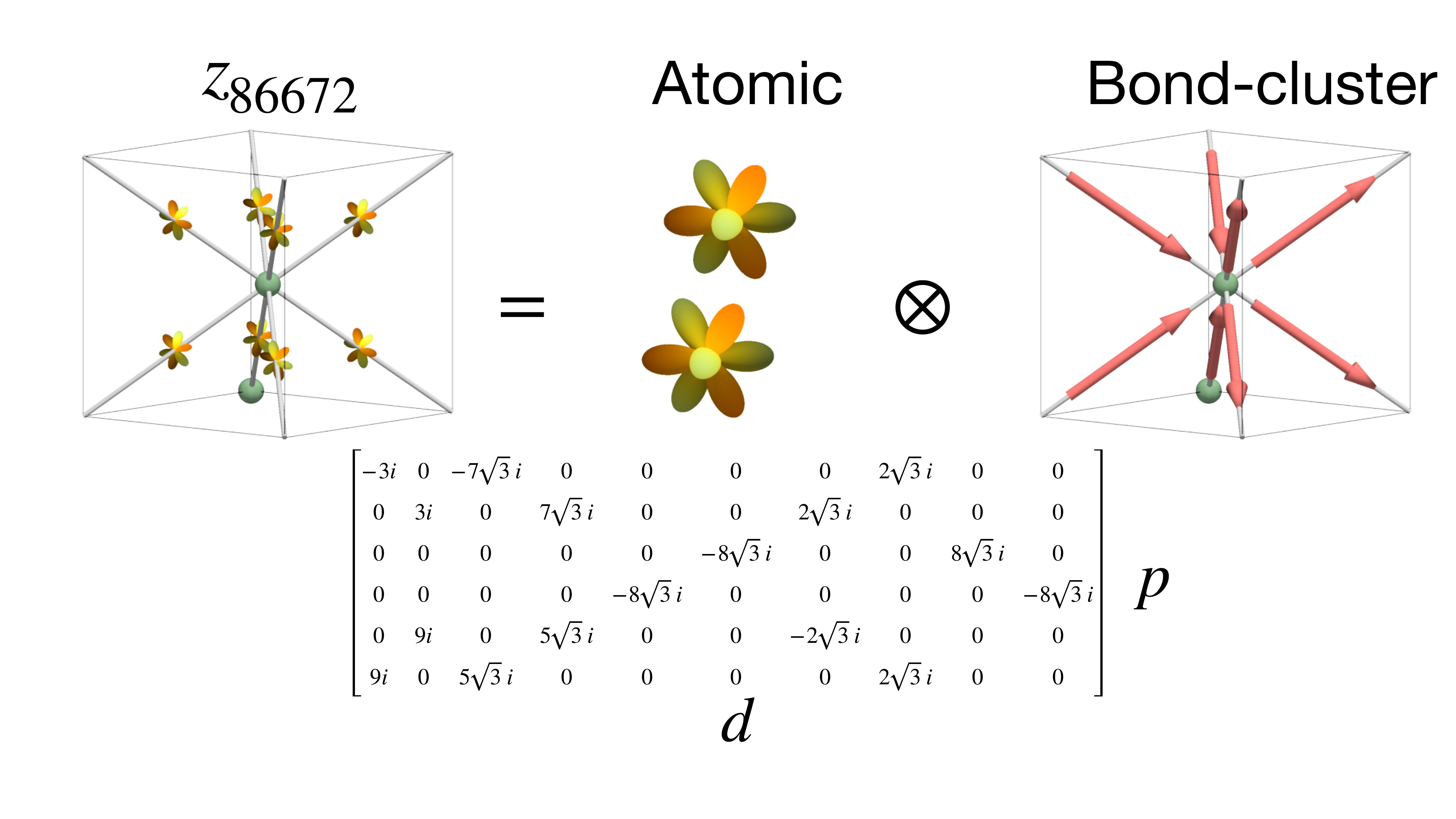}
  \caption{Schematic view of SAMB $z_{86672}$
    ($\mathbb{T}(4,\mathrm{T_{1g}},,0|1,0)$).
    The atomic electric octupole $\mathbb{Q}$ couples with the bond-cluster
    magnetic-toroidal dipole $\mathbb{T}$, corresponding to purely imaginary
  nearest-neighbor hopping.}
  \label{fig:MP86672}
\end{figure}

\begin{figure}[H]
  \centering
  \includegraphics[width=0.8\linewidth]{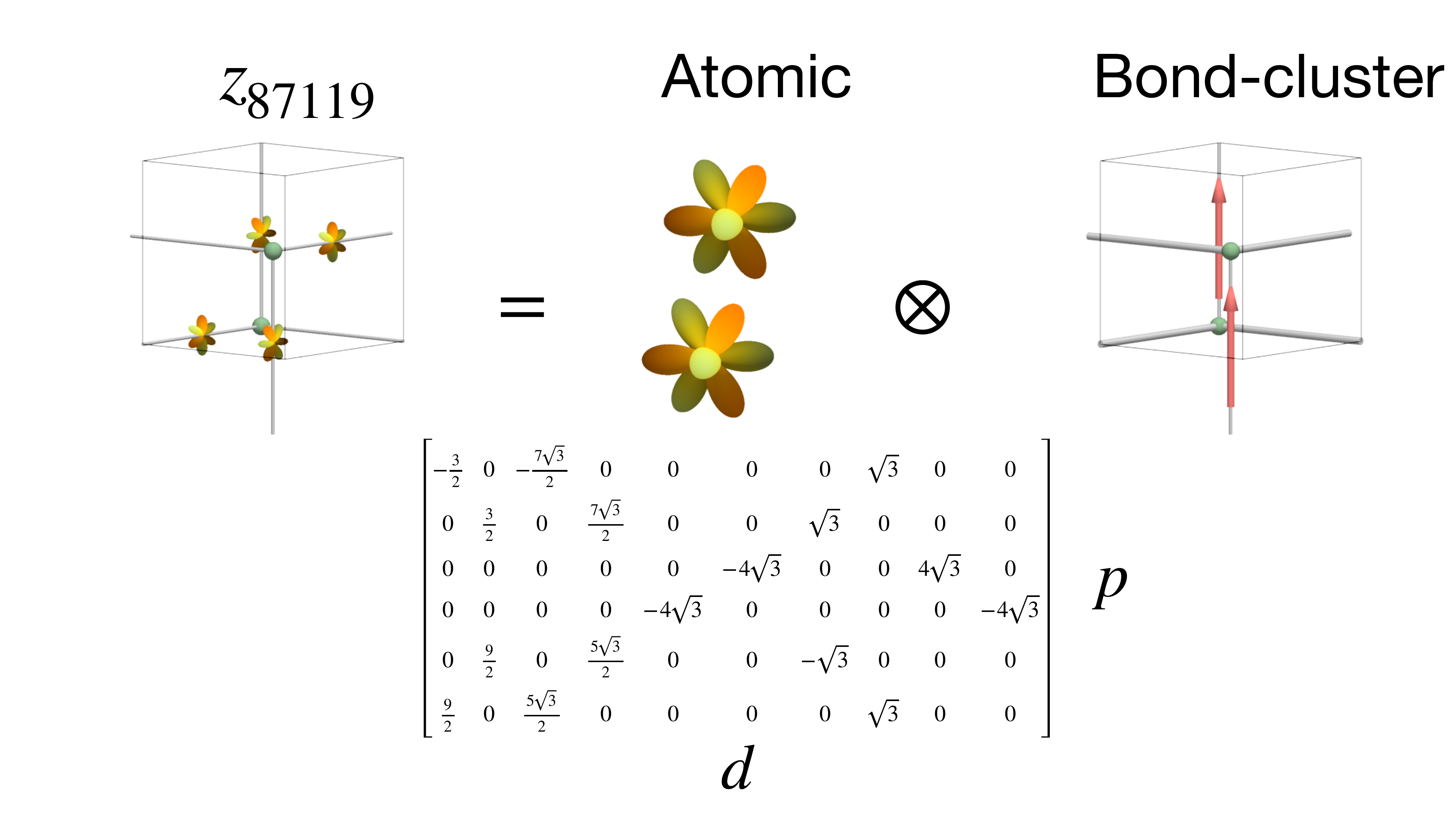}
  \caption{Schematic view of SAMB $z_{87119}$
    ($\mathbb{T}(4,\mathrm{T_{1g}},,0|1,0)$).
    The atomic electric octupole $\mathbb{Q}$ couples with the bond-cluster
    magnetic-toroidal dipole $\mathbb{T}$, corresponding to purely imaginary
  second-nearest-neighbor hopping.}
  \label{fig:MP87119}
\end{figure}

Consistent with this interpretation,
Fig.~\ref{fig:zcoeff_wo_q} shows sizable magnetic-toroidal
multipole $\mathbb{T}$ among the largest non-electric coefficients.
For example, magnetic-toroidal 16-pole $z_{86672}$
($\mathbb{T}(4,\mathrm{T_{1g}},,0|1,0)$) can be written as
\begin{align}
  \mathbb{T}(4,\mathrm{T_{1g}},,0|1,0) =
  & \frac{\sqrt{5}}{4}\,
  \mathbb{Q}^{(\mathrm{a})}(3,\mathrm{T_{1u}},,1|1,0) \otimes
  \mathbb{T}^{(\mathrm{b})}(1,\mathrm{T_{1u}},,2) \nonumber   \\
  & - \frac{\sqrt{5}}{4}\,
  \mathbb{Q}^{(\mathrm{a})}(3,\mathrm{T_{1u}},,2|1,0) \otimes
  \mathbb{T}^{(\mathrm{b})}(1,\mathrm{T_{1u}},,1) \nonumber \\
  & - \frac{\sqrt{3}}{4}\,
  \mathbb{Q}^{(\mathrm{a})}(3,\mathrm{T_{2u}},,1|1,0) \otimes
  \mathbb{T}^{(\mathrm{b})}(1,\mathrm{T_{1u}},,2) \nonumber \\
  & - \frac{\sqrt{3}}{4}\,
  \mathbb{Q}^{(\mathrm{a})}(3,\mathrm{T_{2u}},,2|1,0) \otimes
  \mathbb{T}^{(\mathrm{b})}(1,\mathrm{T_{1u}},,1). \label{eq:MP86672}
\end{align}
Similarly, the atomic electric octupole $\mathbb{Q}$ represents $p$-$d$ order,
while the bond-cluster magnetic-toroidal dipole $\mathbb{T}$
corresponds to purely imaginary
nearest-neighbor hopping (Fig.~\ref{fig:MP86672}).
Beyond the nearest-neighbor terms, second-nearest-neighbor
contributions are also evident in Fig.~\ref{fig:zcoeff_wo_q}.
For example, $z_{87119}$ ($\mathbb{T}(4,\mathrm{T_{1g}},,0|1,0)$)
factorizes in the same way as in Eq.~(\ref{eq:MP86672}), with the bond-cluster
component corresponding to magnetic-toroidal hopping on second-nearest-neighbor
bonds (Fig.~\ref{fig:MP87119}).

\subsection{AHC angular dependence by rotation of magnetic
and magnetic toroidal multipoles}

\subsubsection{Magnetization rotation and AHC angular dependence in
$(111)$ plane}
\begin{figure}[H]
  \begin{subfigure}[b]{0.32\linewidth}
    \centering
    \includegraphics[width=\linewidth]{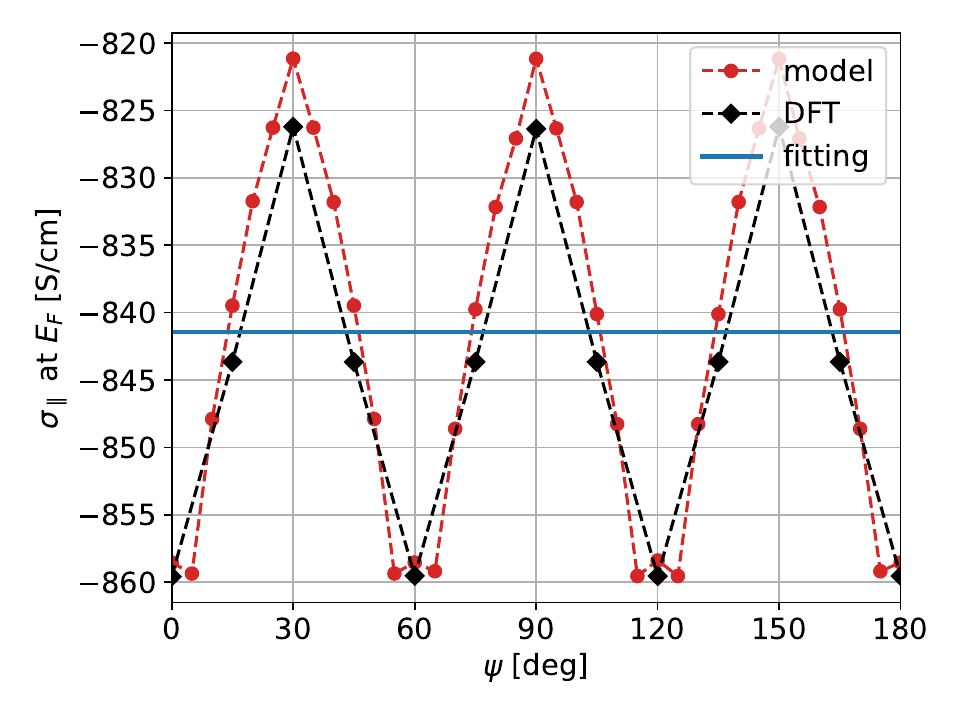}
    \caption{$\sigma_\parallel$}
    \label{fig:ahc_111_para}
  \end{subfigure}\hfill
  \begin{subfigure}[b]{0.32\linewidth}
    \centering
    \includegraphics[width=\linewidth]{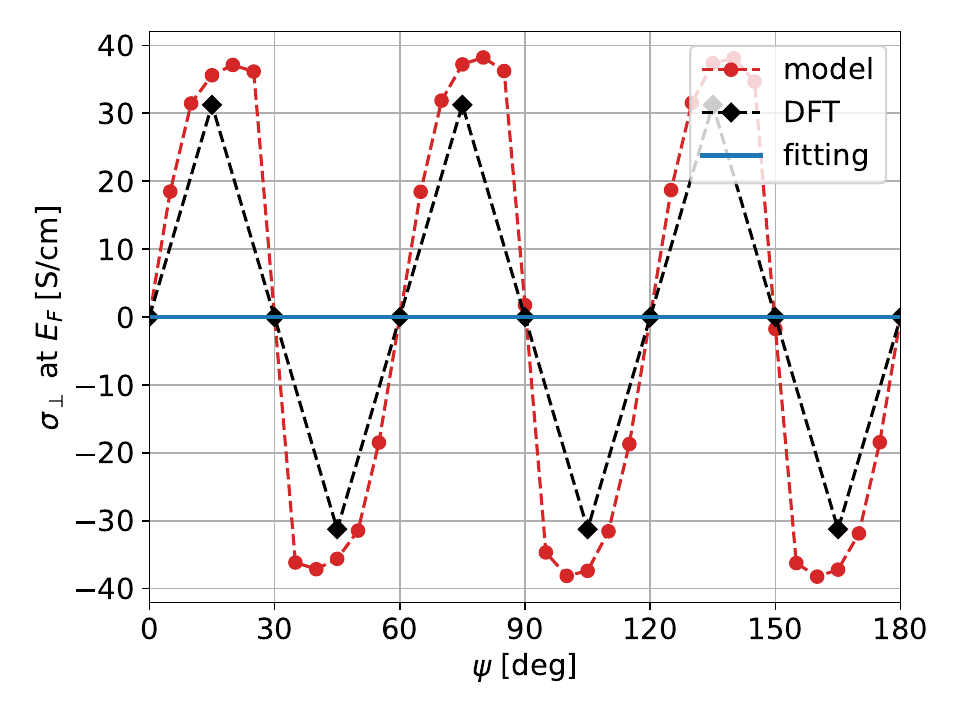}
    \caption{$\sigma_\perp$}
    \label{fig:ahc_111_perp}
  \end{subfigure}\hfill
  \begin{subfigure}[b]{0.32\linewidth}
    \centering
    \includegraphics[width=\linewidth]{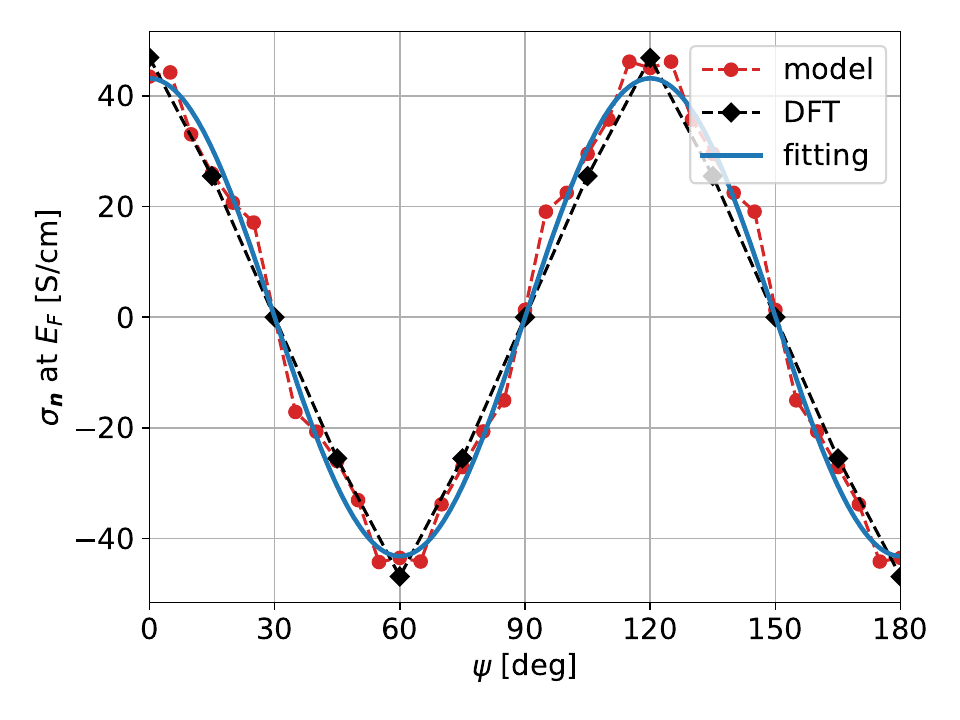}
    \caption{$\sigma_{\bm{n}}$}
    \label{fig:ahc_111_axis}
  \end{subfigure}
  \caption{Angular dependence of AHC components
    ($\sigma_\parallel$, $\sigma_\perp$, $\sigma_{\bm n}$) during
    magnetization rotation in the $(111)$ plane. Symbols denote the
    results from DFT and TRS-Wannier calculations, while solid lines
    represent the fitted theoretical formulas
    ($\sigma_\parallel = \alpha + \tfrac{\beta}{2}$,
      $\sigma_\perp = 0$,
    $\sigma_{\bm{n}} = \tfrac{\sqrt{6}}{6}\beta \cos 3\psi$),
    with fitting parameters
    $\alpha = -894.308$ and $\beta = 105.758\,[\mathrm{S/cm}]$.
  }
  \label{fig:ahc_111}
\end{figure}

Figure~\ref{fig:ahc_111} shows the angular dependence of the AHC
components, $\sigma_\parallel$, $\sigma_\perp$, and $\sigma_{\bm n}$,
during magnetization rotation in the $(111)$ plane. Here, the AHC
values are obtained from two approaches: (i) direct Wannierization at
each magnetization angle (DFT-based) and (ii) the multipole rotation
of the TRS-Wannier Hamiltonian (model-based). The theoretical angular
dependencies given in
Eqs.~(\ref{eq:sigma_parallel_111}-\ref{eq:sigma_n_111}) are also plotted,
where the parameters are fitted using the model-based results.
The fitted values are
$\alpha=-894.308$ and $\beta=105.758 ~[\mathrm{S/cm}]$.

First, we find that the DFT and model-based results are in good
agreement, indicating that the TRS-Wannier method successfully
reproduces the angle-dependent Hamiltonian and can be applied to the
AHC calculation.
Regarding the angular dependence, the out-of-plane component $\sigma_{\bm n}$
(Fig.\ref{fig:ahc_111}(\subref{fig:ahc_111_axis})) reproduces the
periodicity of the
theoretical formula $\sigma_n = \frac{\sqrt{6}}{6}\,\beta \cos 3\psi$.
However, parallel component $\sigma_\parallel$
(Fig.\ref{fig:ahc_111}(\subref{fig:ahc_111_para})) and perpendicular
component $\sigma_\perp$
(Fig.\ref{fig:ahc_111}(\subref{fig:ahc_111_perp})) show significant
deviations from theoretical formulas with discernible 6$\psi$
oscillation periods.
Such $6\psi$ oscillations are clearly expected from the crystal
symmetry and should be explained by higher-order terms that are
neglected in Eq.~(\ref{eq:sigma_cubic}).

\subsubsection{Magnetization rotation and AHC angular dependence in
$(103)$ plane}
\begin{figure}[H]
  \begin{subfigure}[b]{0.32\linewidth}
    \centering
    \includegraphics[width=\linewidth]{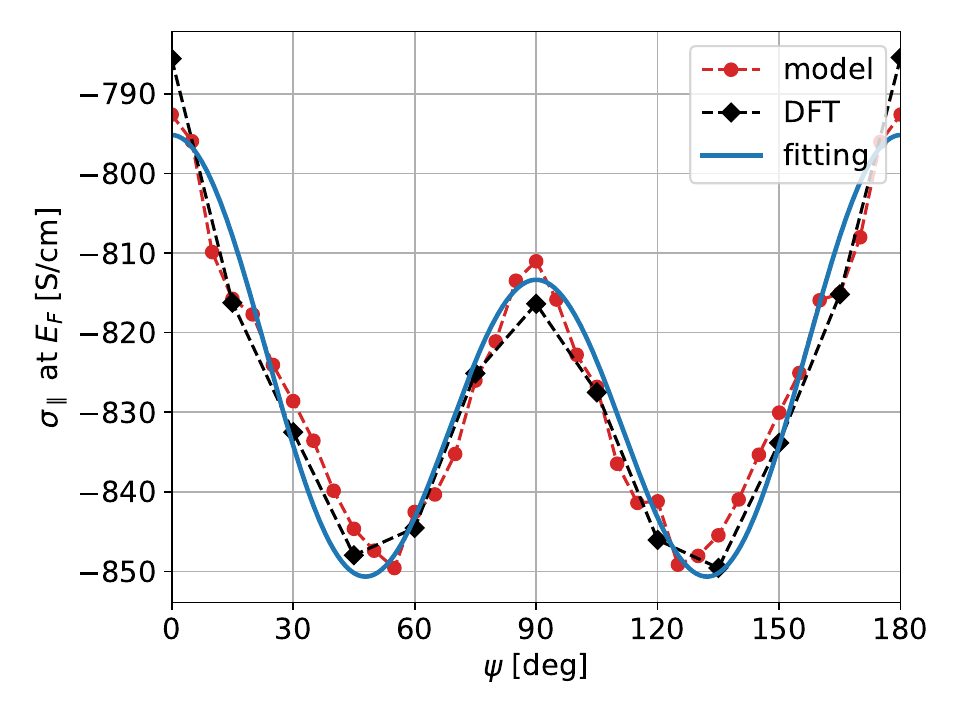}
    \caption{$\sigma_\parallel$}
    \label{fig:ahc_103_para}
  \end{subfigure}\hfill
  \begin{subfigure}[b]{0.32\linewidth}
    \centering
    \includegraphics[width=\linewidth]{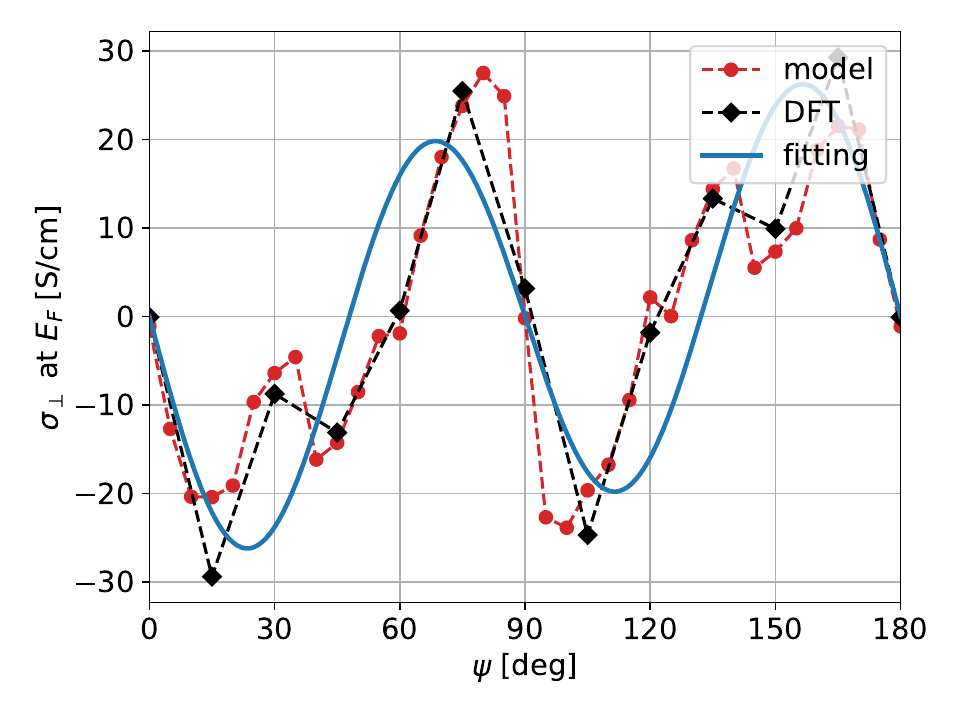}
    \caption{$\sigma_\perp$}
    \label{fig:ahc_103_perp}
  \end{subfigure}\hfill
  \begin{subfigure}[b]{0.32\linewidth}
    \centering
    \includegraphics[width=\linewidth]{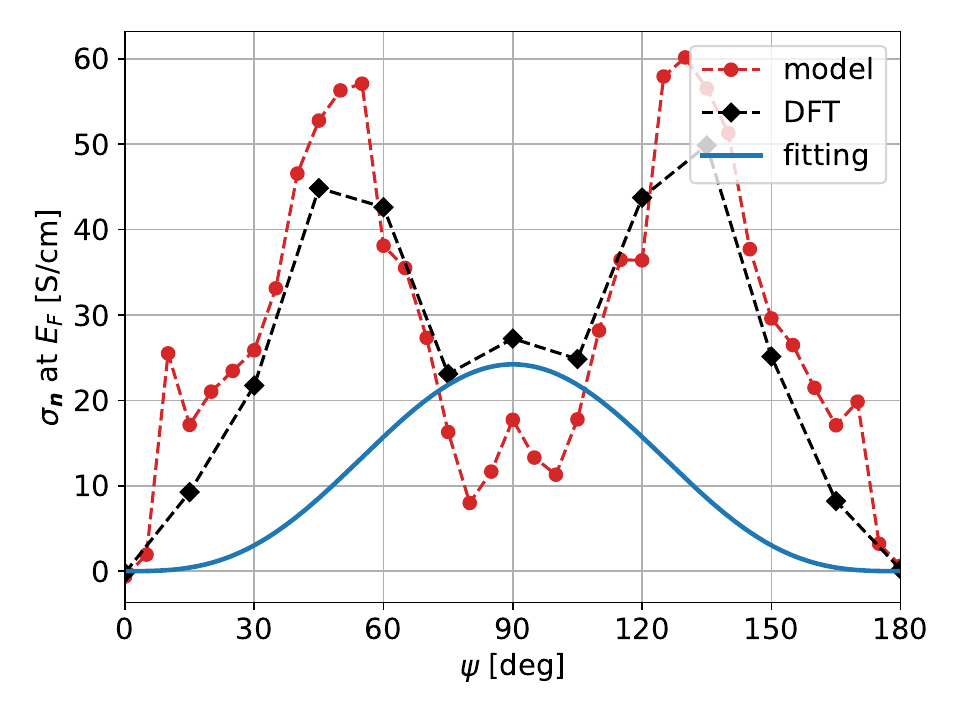}
    \caption{$\sigma_{\bm{n}}$}
    \label{fig:ahc_103_axis}
  \end{subfigure}
  \caption{Angular dependence of AHC components
    ($\sigma_\parallel$, $\sigma_\perp$, $\sigma_{\bm n}$) during
    magnetization rotation in the $(103)$ plane. Symbols denote the
    results from DFT and TRS-Wannier calculations, while solid lines
    represent the fitted theoretical formulas
    ($\sigma_\parallel = \alpha + \beta(\tfrac{273}{400} +
      \tfrac{9}{100}\cos 2\psi + \tfrac{91}{400}\cos 4\psi)$,
      $\sigma_\perp = -\tfrac{\beta}{400}(18\sin 2\psi + 91\sin 4\psi)$,
    $\sigma_{\bm n} = \tfrac{6}{25}\beta \sin^3\psi$),
    with fitting parameters
    $\alpha = -896.116$ and $\beta = 100.928\,[\mathrm{S/cm}]$.
    $\sigma_\parallel$ and $\sigma_\perp$ agree well in periods and phases,
    whereas $\sigma_{\bm n}$ deviates around $\psi=90^{\circ}$,
  indicating higher-order contributions.}
  \label{fig:ahc_103}
\end{figure}

Figure~\ref{fig:ahc_103} shows the angular dependence of the AHC components,
$\sigma_\parallel$, $\sigma_\perp$, and $\sigma_{\bm n}$, during magnetization
rotation in the $(103)$ plane. As in the $(111)$ case, symbols denote the
results from DFT-based Wannierization at each magnetization angle and from the
TRS-Wannier multipole-rotation model, while solid lines represent the
theoretical
formulas in Eqs.~(\ref{eq:sigma_parallel_103}-\ref{eq:sigma_n_103}),
with parameters fitted to the model-based results.
The fitted values are $\alpha=-896.116$ and
$\beta=100.928 ~[\mathrm{S/cm}]$.

For the parallel component $\sigma_\parallel$
(Fig.\ref{fig:ahc_103}(\subref{fig:ahc_103_para})) and perpendicular
component $\sigma_\perp$
(Fig.\ref{fig:ahc_103}(\subref{fig:ahc_103_perp})), periods and phases
nearly match theoretical formula $\sigma_\parallel =
\alpha + \beta\left( \frac{273}{400} + \frac{9}{100}\cos 2\psi +
\frac{91}{400}\cos 4\psi \right)$ and $\sigma_\perp =
-\frac{\beta}{400}\bigl(18 \sin 2\psi + 91 \sin 4\psi\bigr)$, showing
the model reproduces main angular dependence features.
Since the fitted values of $\alpha$ and $\beta$ are consistent between the
$(111)$ plane case and the present case, the low-order parameters are
robust against surface orientations.
However, in this case, the out-of-plane component $\sigma_{\bm n}$
(Fig.\ref{fig:ahc_103}(\subref{fig:ahc_103_axis})) shows
two local minima at $\phi = 0^{\circ}$ and $90^{\circ}$ which cannot
be reproduced by the
theoretical formula, $\sigma_n =\frac{6}{25}\beta \sin^3\psi$.
A similar valley-like feature near
$\psi = 90^\circ$ has also been suggested by an experiment \cite{Peng2024-oj}.
These deviations indicate that a low-order expansion in magnetization
angle is insufficient for the $(103)$ plane as well.
We address higher-order and magnetic-toroidal contributions to the
AHC in the next section.

\subsection{Rank-resolved multipole and strain effects on IAHE in the
(103) plane}

\subsubsection{Rank-wise contribution analysis in $(103)$ plane}
\begin{figure}[H]
  \centering
  \begin{subfigure}[b]{0.31\linewidth}
    \centering
    \includegraphics[width=\linewidth]{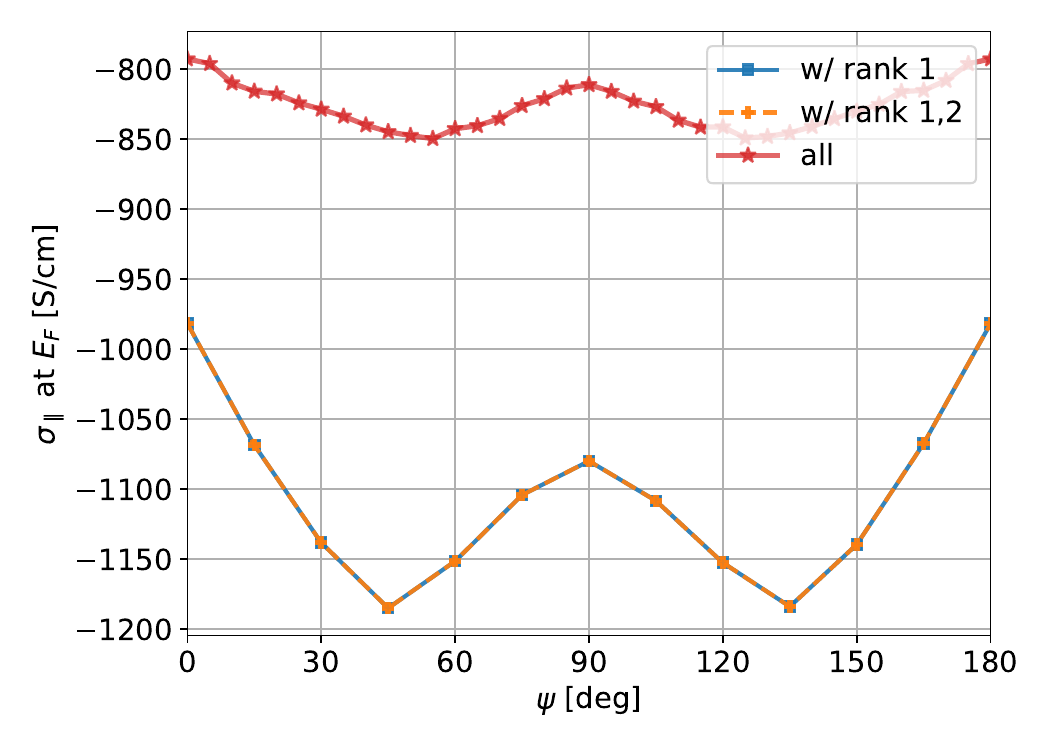}
    \caption{$\sigma_\parallel$}
    \label{fig:rank103_group1_para}
  \end{subfigure}\hfill
  \begin{subfigure}[b]{0.31\linewidth}
    \centering
    \includegraphics[width=\linewidth]{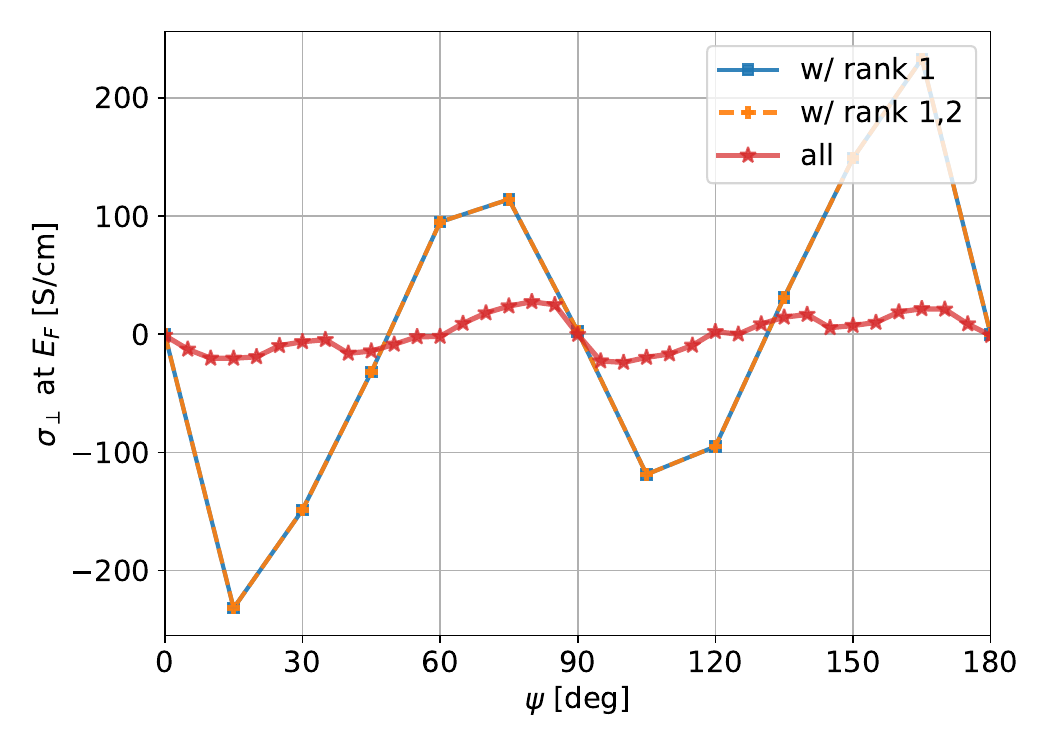}
    \caption{$\sigma_\perp$}
    \label{fig:rank103_group1_perp}
  \end{subfigure}\hfill
  \begin{subfigure}[b]{0.31\linewidth}
    \centering
    \includegraphics[width=\linewidth]{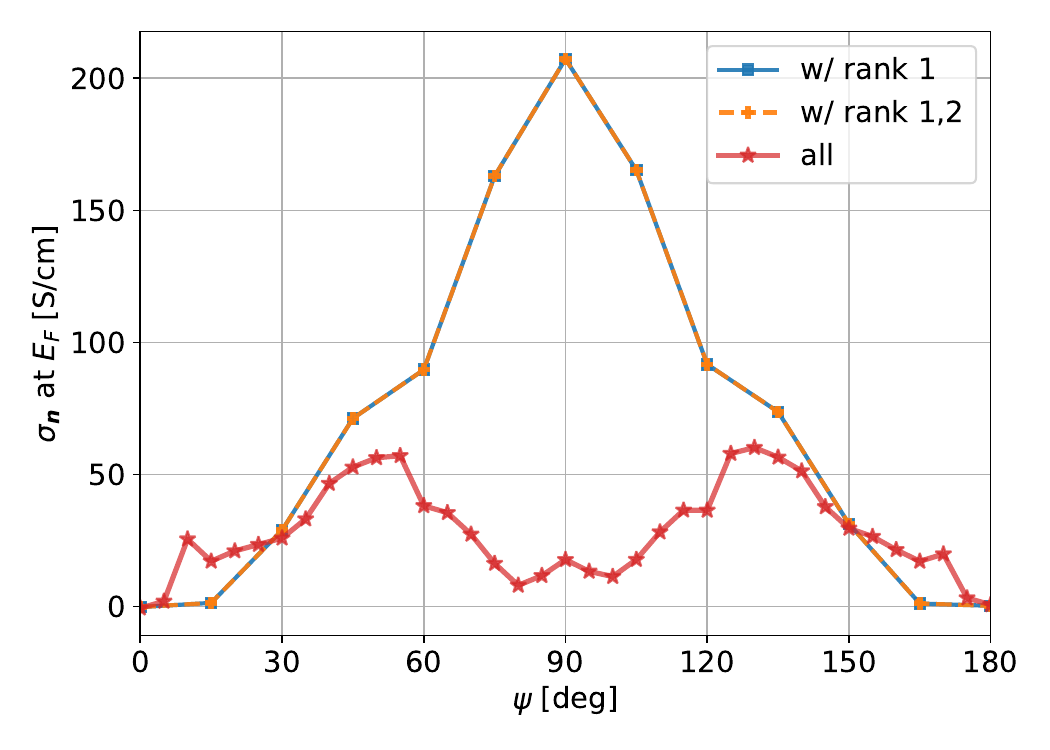}
    \caption{$\sigma_{\bm{n}}$}
    \label{fig:rank103_group1_axis}
  \end{subfigure}

  \par\medskip

  \begin{subfigure}[b]{0.31\linewidth}
    \centering
    \includegraphics[width=\linewidth]{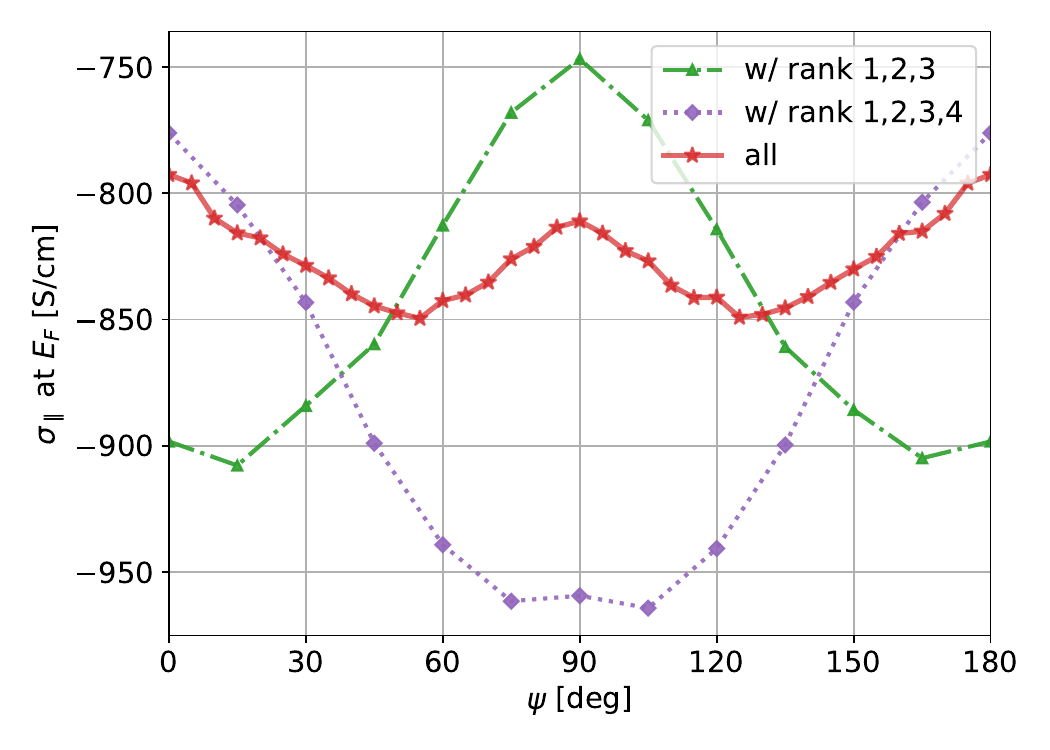}
    \caption{$\sigma_\parallel$}
    \label{fig:rank103_group2_para}
  \end{subfigure}\hfill
  \begin{subfigure}[b]{0.31\linewidth}
    \centering
    \includegraphics[width=\linewidth]{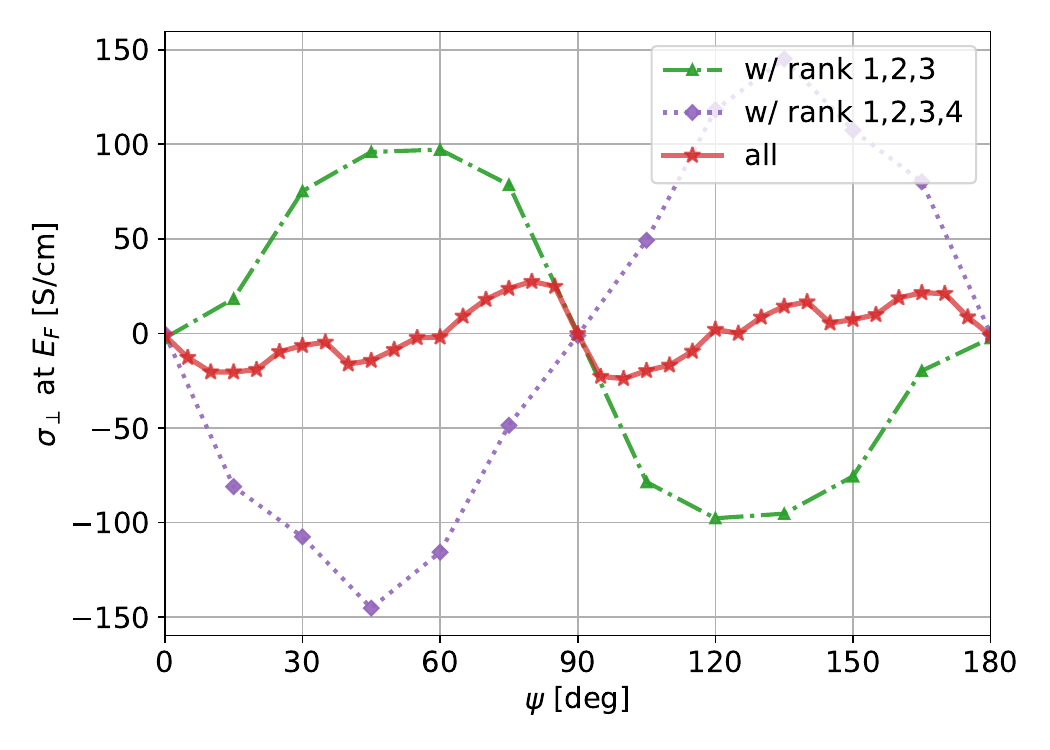}
    \caption{$\sigma_\perp$}
    \label{fig:rank103_group2_perp}
  \end{subfigure}\hfill
  \begin{subfigure}[b]{0.31\linewidth}
    \centering
    \includegraphics[width=\linewidth]{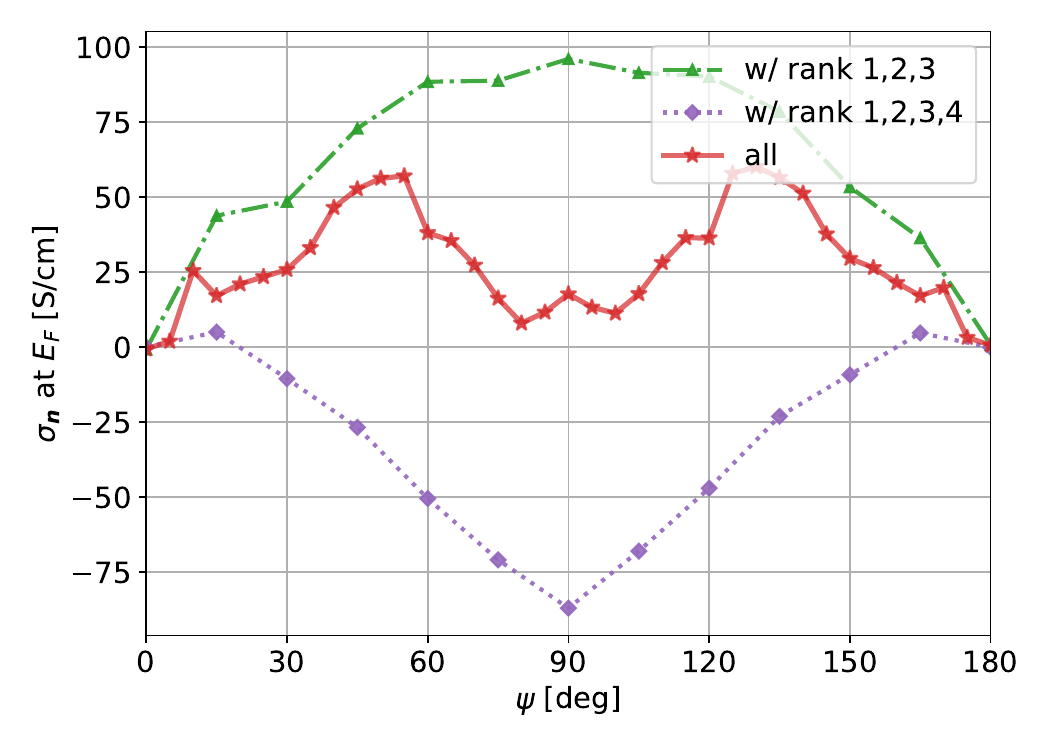}
    \caption{$\sigma_{\bm{n}}$}
    \label{fig:rank103_group2_axis}
  \end{subfigure}

  \par\medskip

  \begin{subfigure}[b]{0.31\linewidth}
    \centering
    \includegraphics[width=\linewidth]{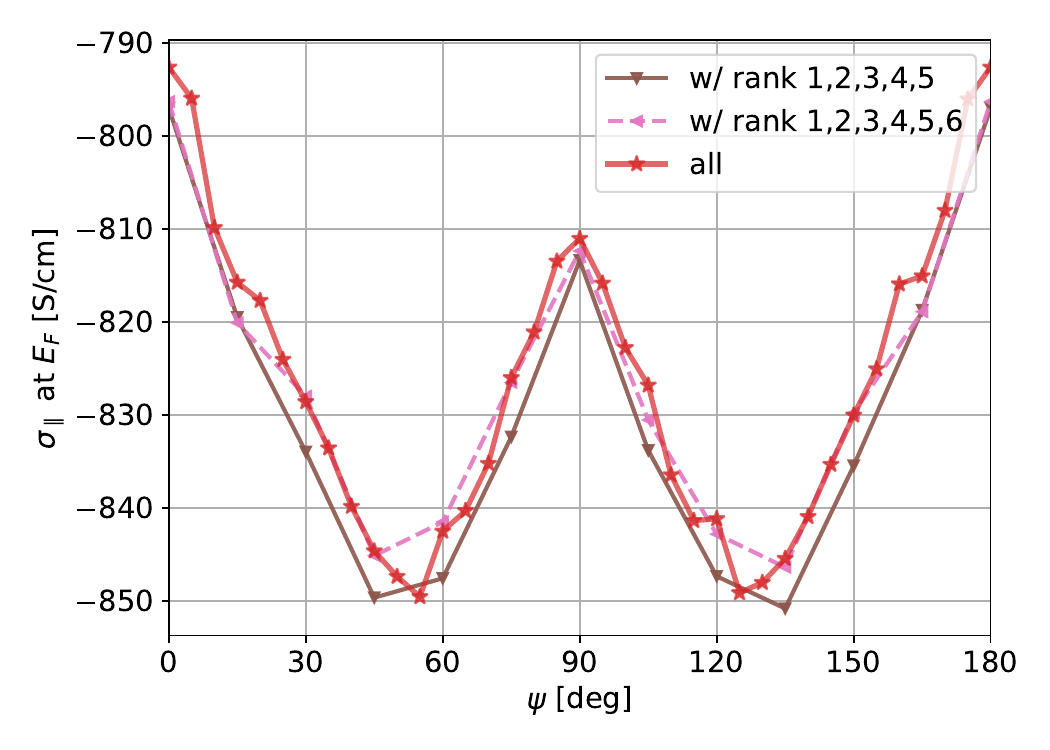}
    \caption{$\sigma_\parallel$}
    \label{fig:rank103_group3_para}
  \end{subfigure}\hfill
  \begin{subfigure}[b]{0.31\linewidth}
    \centering
    \includegraphics[width=\linewidth]{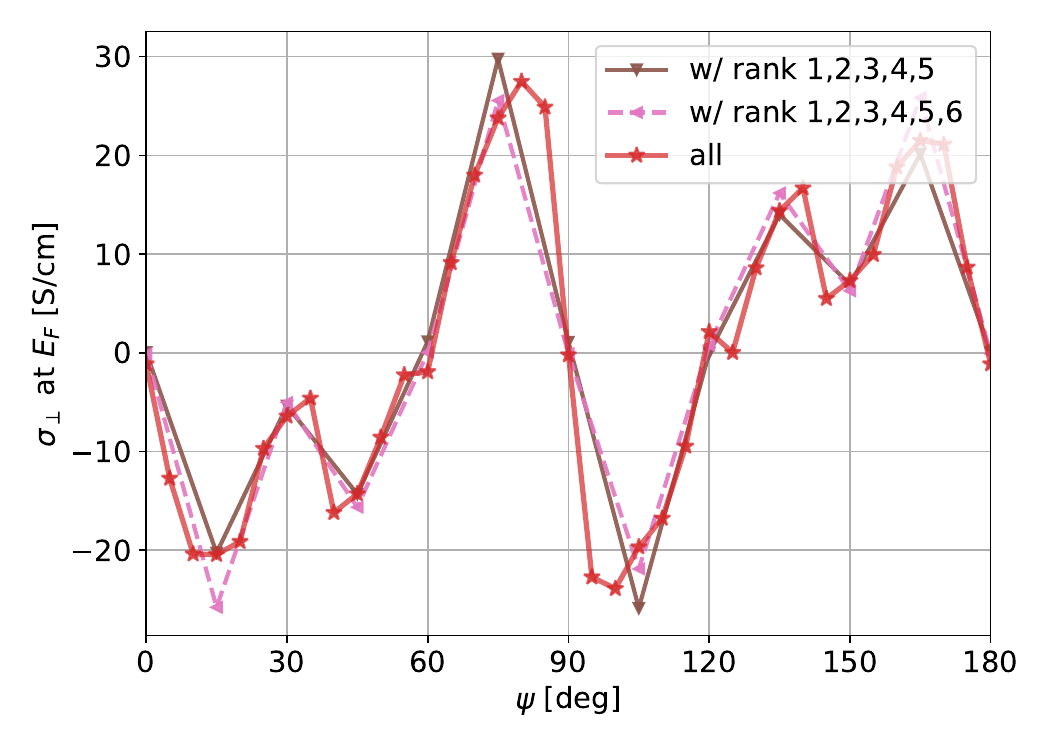}
    \caption{$\sigma_\perp$}
    \label{fig:rank103_group3_perp}
  \end{subfigure}\hfill
  \begin{subfigure}[b]{0.31\linewidth}
    \centering
    \includegraphics[width=\linewidth]{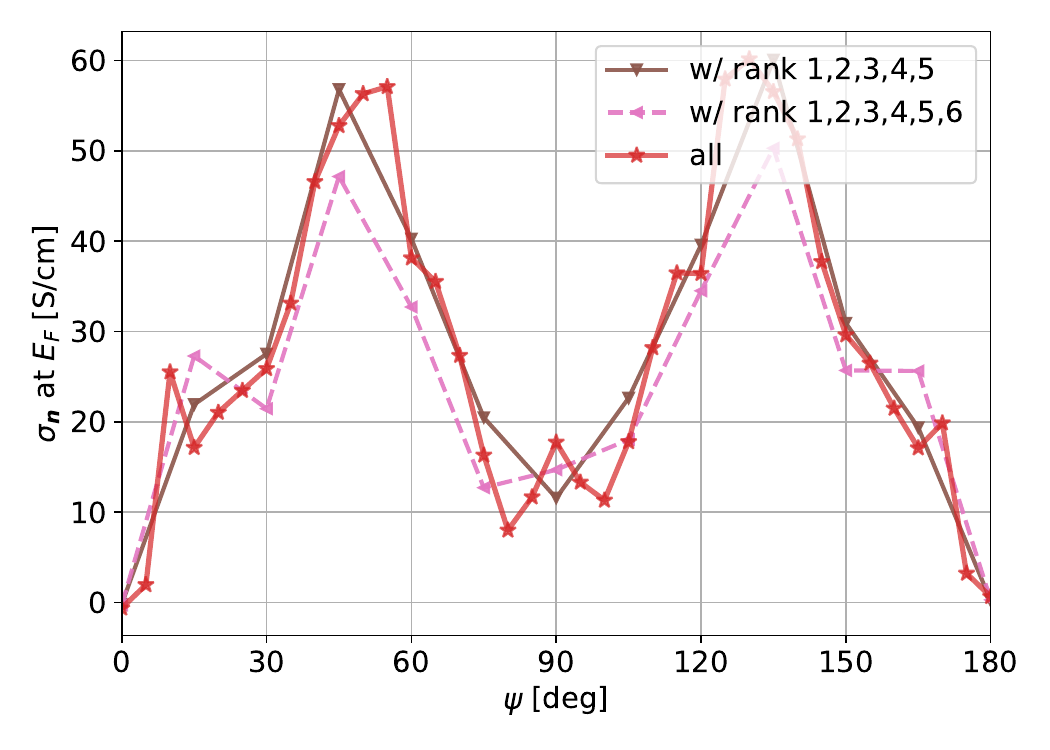}
    \caption{$\sigma_{\bm{n}}$}
    \label{fig:rank103_group3_axis}
  \end{subfigure}

  \caption{Angular dependence of $\sigma_\parallel$ (a,d,g),
    $\sigma_\perp$ (b,e,h) and $\sigma_{\bm{n}}$ (c,f,i) for
    different truncations of
    the multipole decomposition: (a-c) results including multipoles up to
    rank 1 and 2,
    (d-f) up to rank 3 and 4,
    (g-i) up to rank 5 and 6.
  Results including all ranks are also shown as references.}
  \label{fig:rank103_all}
\end{figure}

Next, by examining the contribution of each term in the multipole
decomposition, we investigate which multipoles play dominant roles.
Figure~\ref{fig:rank103_all} summarizes the angle dependence of
$\sigma_\parallel$, $\sigma_\perp$, and $\sigma_{\bm n}$, for
different truncations of the multipole decomposition.
Considering only the lowest rank magnetic dipole $\mathbb{M}_1$,
the parallel component $\sigma_\parallel$
(Fig.\ref{fig:rank103_all}(\subref{fig:rank103_group1_para})) and
perpendicular component $\sigma_\perp$
(Fig.\ref{fig:rank103_all}(\subref{fig:rank103_group1_perp})) well
reproduce basic angular dependence of theoretical formulas given in
Eqs.~(\ref{eq:sigma_parallel_103}-\ref{eq:sigma_n_103}).
Particularly, the out-of-plane component $\sigma_{\bm n}$
(Fig.\ref{fig:rank103_all}(\subref{fig:rank103_group1_axis})) shows
a $\sin^3\psi$ shape with clear peak at $\psi=90^{\circ}$.
The $\mathbb{M}_1$-only case has better agreement with theoretical
formulas, showing that $\mathbb{M}_1$ has dominant contributions
determining overall shape.
The $\mathbb{T}_2$ component is negligible and does not contribute to AHC.

When adding more ranks, higher-order components like $\mathbb{M}_3$,
$\mathbb{T}_4$, and $\mathbb{M}_5$ each have amplitudes comparable to
$\mathbb{M}_1$ and significantly modulate the angular dependence.
As shown in
Fig.\ref{fig:rank103_all}(\subref{fig:rank103_group2_para}-\subref{fig:rank103_group2_axis}),
adding $\mathbb{M}_3$ and
$\mathbb{T}_4$ introduces new harmonic components to
$\sigma_\parallel$ and $\sigma_\perp$, and in the subsequent
Fig.\ref{fig:rank103_all}(\subref{fig:rank103_group3_para}-\subref{fig:rank103_group3_axis}),
adding $\mathbb{M}_5$ and
$\mathbb{T}_6$ contributions causes these harmonic components to
cancel each other, finally converging to the overall shape.
This cancellation effect reduces amplitudes compared to the
$\mathbb{M}_1$ alone case, while causing fine deviations from theoretical
curves for $\sigma_{\bm n}$ and changes in peak shapes around $\psi=90^{\circ}$.
Especially in
Fig.~\ref{fig:rank103_all}(\subref{fig:rank103_group2_para}-\subref{fig:rank103_group2_axis}),
we identify the rank-4 magnetic-toroidal component $\mathbb{T}_4$ as
providing a large contribution with the sign opposite to the others.

\begin{figure}[H]
  \begin{subfigure}[b]{0.32\linewidth}
    \centering
    \includegraphics[width=\linewidth]{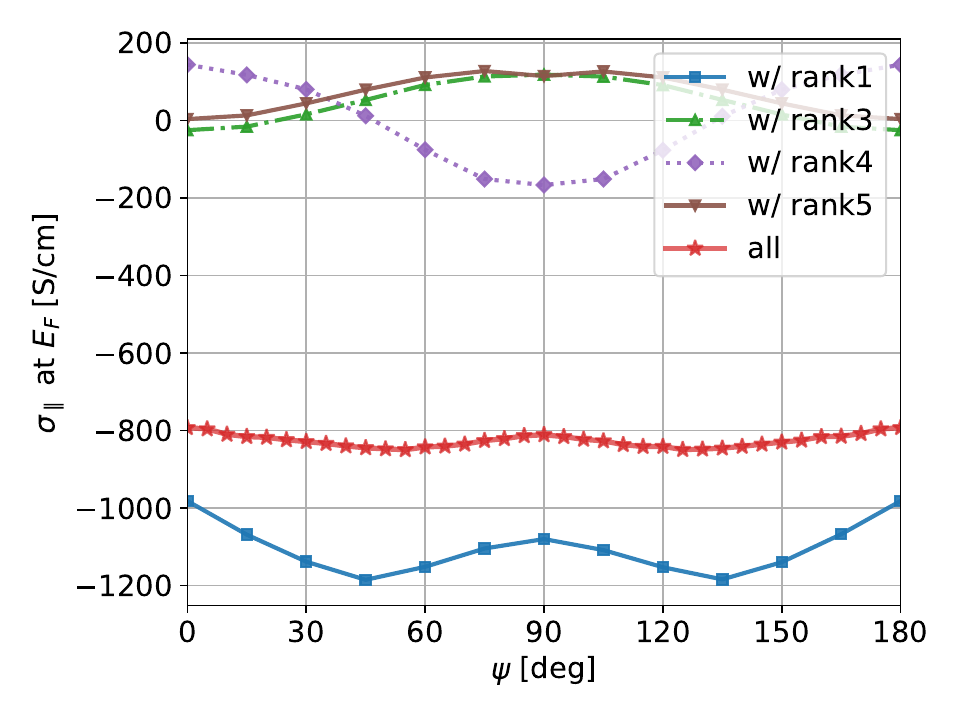}
    \caption{$\sigma_\parallel$}
    \label{fig:rank103_single_para}
  \end{subfigure}\hfill
  \begin{subfigure}[b]{0.32\linewidth}
    \centering
    \includegraphics[width=\linewidth]{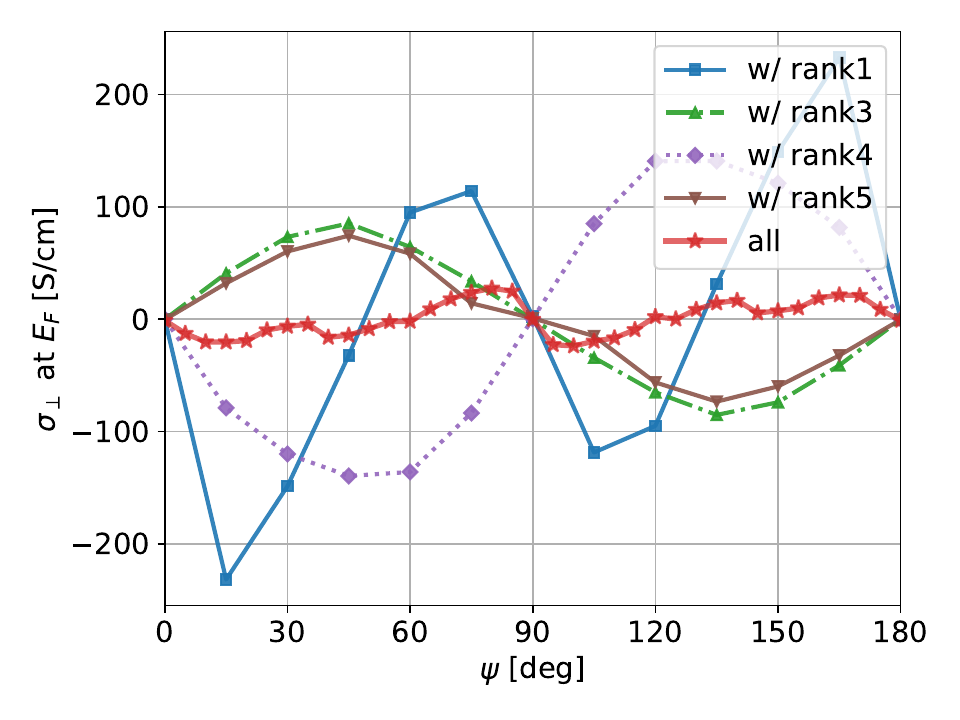}
    \caption{$\sigma_\perp$}
    \label{fig:rank103_single_perp}
  \end{subfigure}\hfill
  \begin{subfigure}[b]{0.32\linewidth}
    \centering
    \includegraphics[width=\linewidth]{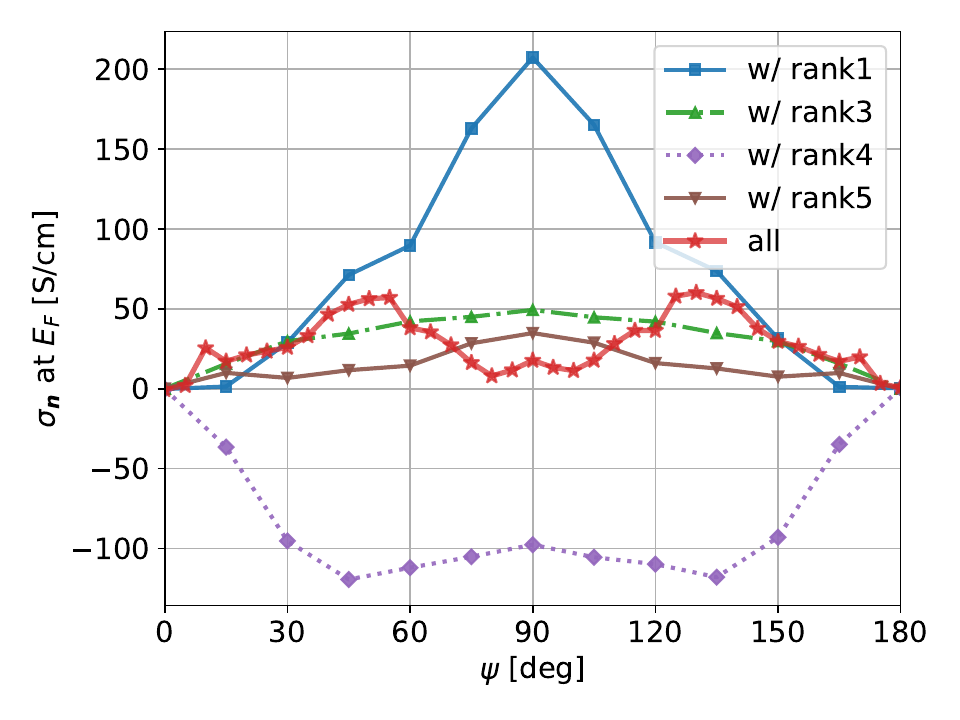}
    \caption{$\sigma_{\bm{n}}$}
    \label{fig:rank103_single_axis}
  \end{subfigure}
  \caption{Angular dependence of AHC components ($\sigma_\parallel$,
    $\sigma_\perp$, $\sigma_{\bm{n}}$) during magnetization rotation in
    the $(103)$ plane, comparing cases where only single ranks such as
    $\mathbb{M}_1$, $\mathbb{M}_3$, $\mathbb{T}_4$, or $\mathbb{M}_5$
    are included.
    Consistent with Fig.~\ref{fig:rank103_all}, the rank-4
    magnetic-toroidal component $\mathbb{T}_4$ contributes to
    the AHC with the opposite
    sign to $\mathbb{M}_1$, $\mathbb{M}_3$, and $\mathbb{M}_5$ and reshapes
  $\sigma_{\bm{n}}$ around $\psi = 90^{\circ}$.}
  \label{fig:rank103_single}
\end{figure}

Figure \ref{fig:rank103_single} shows the angular dependence obtained
by considering only one of $\mathbb{M}_{3}$, $\mathbb{T}_{4}$, or
$\mathbb{M}_{5}$ as the terms corresponding to magnetization, $H^a$.
Consistent with the cumulative analysis in Fig.~\ref{fig:rank103_all},
single-rank rotations of $\mathbb{M}_3$, $\mathbb{T}_4$, and
$\mathbb{M}_5$ reproduce similar trends, such that $\mathbb{T}_4$
contributes with the opposite sign and reshapes the valley of
$\sigma_{\bm{n}}$ near $\psi = 90^{\circ}$
(Fig.~\ref{fig:rank103_single}(\subref{fig:rank103_single_axis})).
The next subsection examines prospects for controlling $\mathbb{T}_4$.

\subsubsection{Strain-induced symmetry lowering and IAHE modulation
in the (103) plane}
\begin{figure}[H]
  \begin{subfigure}{0.48\linewidth}
    \centering
    \includegraphics[width=0.8\linewidth]{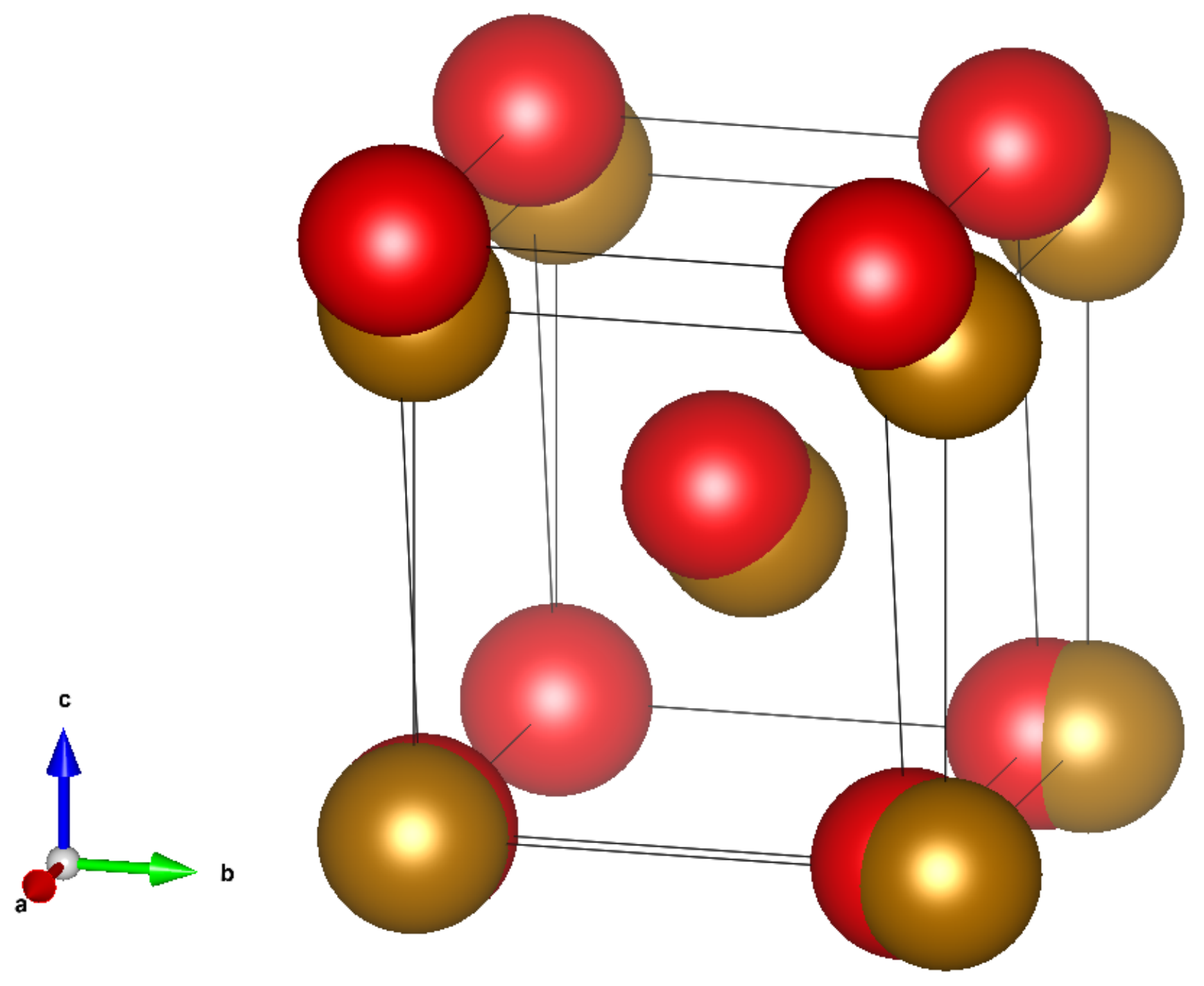}
    \caption{Positive strain along [103].}
    \label{fig:structure_p20}
  \end{subfigure}
  \begin{subfigure}{0.48\linewidth}
    \centering
    \includegraphics[width=0.8\linewidth]{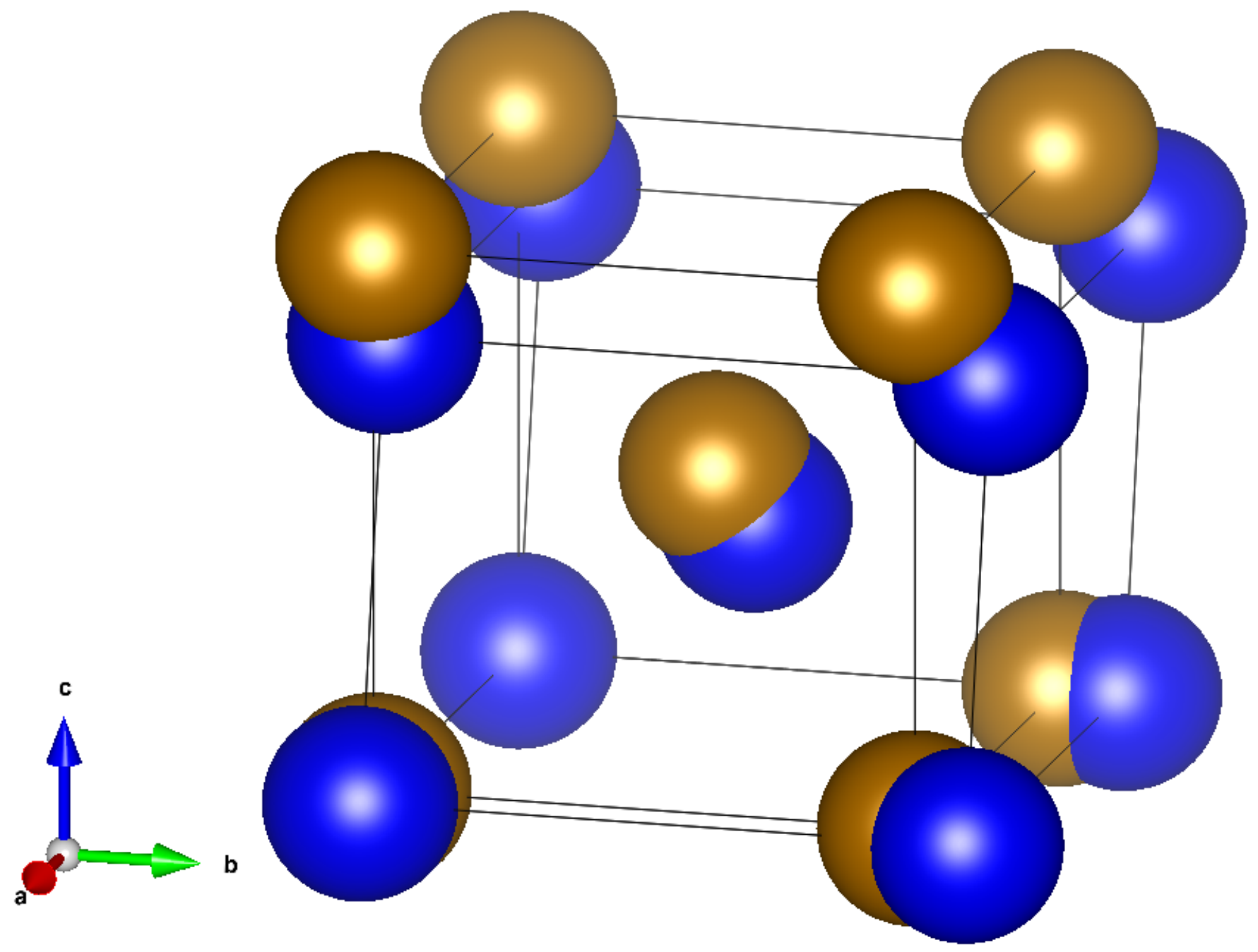}
    \caption{Negative strain along [103].}
    \label{fig:structure_m20}
  \end{subfigure}
  \caption{Crystal structures of body-centered cubic iron (bcc
    $\ce{Fe}$) with positive strain and with negative
  strain along the [103] direction.}
  \label{fig:Fe_structures}
\end{figure}

We consider a body-centered cubic crystal subjected to a
uniform, volume-conserving strain along the [103] direction
(Fig.~\ref{fig:Fe_structures}). Let $\bm{a}_i$ denote the original
lattice vectors and $\bm{a}'_i$ the deformed ones. They are related by
the deformation gradient $F$ as
\begin{align}
  \bm{a}'_i &= F \bm{a}_i,\\
  F &= \lambda^{-1/2} I + \bigl(\lambda - \lambda^{-1/2}\bigr)
  (\bm{n} \otimes \bm{n}),
\end{align}
where $I$ is the identity tensor, $\lambda$ is the stretch ratio, and
$\bm{n}$ is a unit vector specifying the strain direction, that is,
$\bm{n}=\frac{1}{\sqrt{10}}(1,0,3)$.

Under this strain, the point group of the system is reduced from
$O_{\mathrm{h}}$ to $C_{\mathrm{2h}}$. Table~\ref{tab:multipoles}
summarizes the correspondence between multipoles and irreducible
representations in $O_{\mathrm{h}}$ and $C_{\mathrm{2h}}$. From the
table, $M_z$, $M_z^\alpha$, and $T_z^\alpha$, which belong to
$\mathrm{T_{1g}}$ in $O_{\mathrm{h}}$, transform as $\mathrm{A_g}$ in
$C_{\mathrm{2h}}$. This reduction indicates that, in $C_{\mathrm{2h}}$,
many other magnetic and magnetic-toroidal multipoles in $\mathrm{A_g}$
become active and may contribute to the AHC.

\begin{table}[H]
  \caption{Multipoles and irreducible representations in
    $O_{\mathrm{h}}$ and $C_{\mathrm{2h}}$ point groups. Excerpted from
  Ref.~\cite{Hayami2018-fn}. \label{tab:multipoles}}
  \centering
  \begin{tabular}{cc@{\hspace{2em}}cc}
    \toprule
    $\mathbb{M}$ & $\mathbb{T}$ & $O_{\mathrm{h}}$ & $C_{\mathrm{2h}}$ \\
    \midrule
    -- &  $T_0$, $T_4$  &  $\mathrm{A_{1g}}$  &  $\mathrm{A_g}$   \\
    $M_{xyz}$ &  --  &  $\mathrm{A_{2g}}$  &   $\mathrm{A_g}$  \\
    -- &  $T_u$, $T_{4u}$  &  $\mathrm{E_g}$  &   $\mathrm{A_g}$  \\
    -- &  $T_v$, $T_{4v}$  &  $\mathrm{E_g}$  &   $\mathrm{A_g}$  \\
    $M_x$, $M^\alpha_x$ &  $T^\alpha_{4x}$  &  $\mathrm{T_{1g}}$  &
    $\mathrm{B_g}$  \\
    $M_y$, $M^\alpha_y$ &  $T^\alpha_{4y}$  &  $\mathrm{T_{1g}}$  &
    $\mathrm{B_g}$  \\
    $M_z$, $M^\alpha_z$ &  $T^\alpha_{4z}$  &  $\mathrm{T_{1g}}$  &
    $\mathrm{A_g}$  \\
    $M^\beta_x$ &  $T_{yz}$, $T^\beta_{4x}$  &  $\mathrm{T_{2g}}$  &
    $\mathrm{B_g}$  \\
    $M^\beta_y$ &  $T_{zx}$, $T^\beta_{4y}$  &  $\mathrm{T_{2g}}$  &
    $\mathrm{B_g}$  \\
    $M^\beta_z$ &  $T_{xy}$, $T^\beta_{4z}$  &  $\mathrm{T_{2g}}$  &
    $\mathrm{A_g}$  \\
    \bottomrule
  \end{tabular}
\end{table}

\begin{figure}[H]
  \centering
  \begin{subfigure}[b]{0.45\linewidth}
    \centering
    \includegraphics[width=\linewidth]{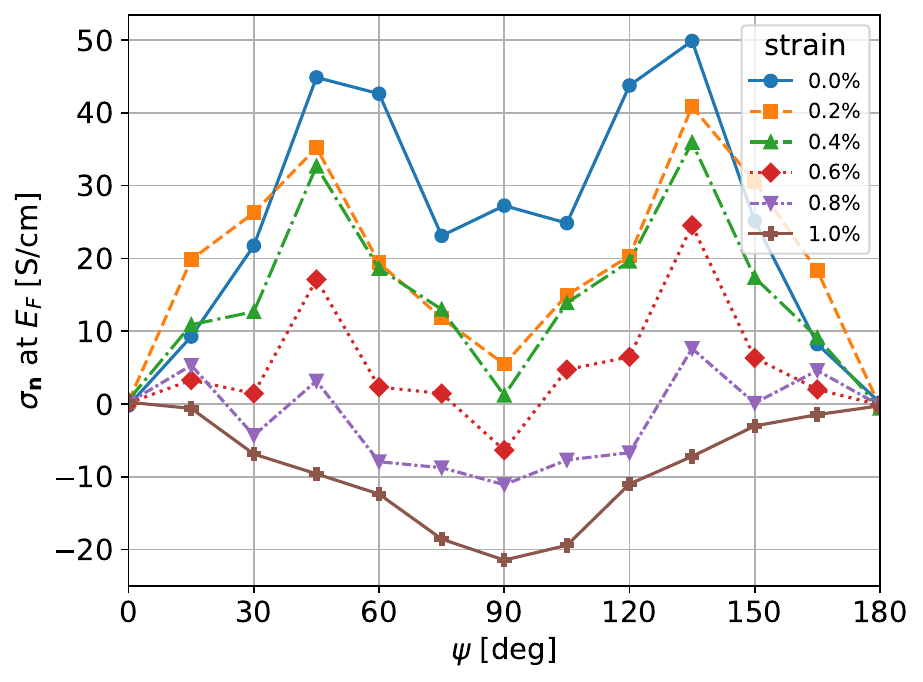}
    \caption{$\sigma_{\bm{n}}$ (positive strain along [103])}
    \label{fig:plus_strain}
  \end{subfigure}\hfill
  \begin{subfigure}[b]{0.45\linewidth}
    \centering
    \includegraphics[width=\linewidth]{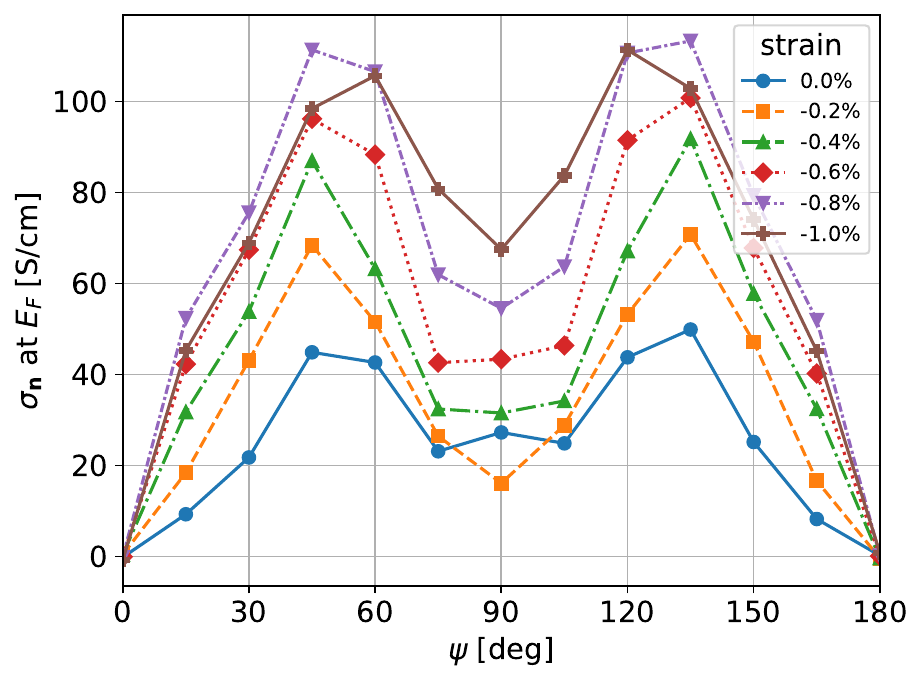}
    \caption{$\sigma_{\bm{n}}$ (negative strain along [103])}
    \label{fig:minus_strain}
  \end{subfigure}
  \caption{Effect of uniaxial strain applied along [103] on the out-of-plane AHC
    component $\sigma_{\bm n}$ during magnetization rotation in the $(103)$
    plane. (a) tensile (positive) strain; (b) compressive (negative)
  strain.}\label{fig:strain_103}
\end{figure}

Figure~\ref{fig:strain_103} shows the angular dependence of the
out-of-plane AHC component $\sigma_{\bm n}$ during magnetization
rotation in the $(103)$ plane, under both tensile (positive) and
compressive (negative) strains along [103]. In both cases, the peak
structure around $\psi=90^{\circ}$ is significantly altered compared to
the unstrained case (Fig.~\ref{fig:ahc_103}(\subref{fig:ahc_103_axis})).
Under compressive strain, the valley-like feature around
$\psi=90^{\circ}$ becomes relatively weaker. In contrast, tensile
strain deepens this valley, and at +1\% strain, the sign of
$\sigma_{\bm n}$ is inverted for all angles $\psi$.
This indicates that strain enhances the weight of $\mathrm{A_g}$
channels such as $\mathbb{T}_4$, which contributes with the opposite
sign, thereby deepening the valley and inverting the sign
of $\sigma_{\bm n}$ over the entire angular range.

\subsection{Minimal two-orbital $p_z$-$d_{xy}$ model Hamiltonian}
To understand the strain effect on the angular dependence of the IAHE,
we propose a minimal model Hamiltonian that reproduces the
valley-like IAHE response in the (103) plane. The Hamiltonian
consists of five terms---on-site energy $H_{\rm on}$, nearest-neighbor
hopping $H_{\rm hop}$, exchange interaction $H_{\rm ex}$, and
magnetic-toroidal hopping on first- and second-nearest-neighbor
bonds, $H^{(1)}_{\rm MT}$ and $H^{(2)}_{\rm MT}$, explicitly written as,
\begin{align}
  &H=H_{\rm on}+H_{\rm hop}+H_{\rm ex}+H^{(1)}_{\rm MT}+H^{(2)}_{\rm MT}, \\
  &H_{\rm on}
  =\sum_i \Psi_i^\dagger\left[
    \varepsilon_0\tau_0\otimes s_0
    +\frac{\Delta_0}{2}\tau_z\otimes s_0
  \right]\Psi_i,\\
  &H_{\rm hop}
  =t_0\sum_{\langle ij\rangle}\Psi_i^\dagger\left[
    \tau_0\otimes s_0
  \right]\Psi_j,\\
  &H_{\rm ex}
  =-\Delta\sum_i \Psi_i^\dagger\left[
    \tau_0\otimes\hat{\bm M}(\psi)\cdot\bm{s}
  \right]\Psi_i,\\
  &H^{(1)}_{\rm MT}
  =-it^{(1)}_{\mathrm{T}}\sum_{\langle ij\rangle}
  \Psi_i^\dagger\left[
    \tau_y \otimes \big(\hat{\bm{M}}
    (\psi)\times\bm{s}\big)\cdot\bm{d}_{ij}
  \right]\Psi_j+{\rm h.c.}, \\
  &H^{(2)}_{\rm MT}
  =-it^{(2)}_{\mathrm{T}}\sum_{\langle\langle ij\rangle\rangle}
  \Psi_i^\dagger\left[
    \tau_y \otimes \big(\hat{\bm{M}}
    (\psi)\times\bm{s}\big)\cdot\bm{d}_{ij}
  \right]\Psi_j+{\rm h.c.}. \label{eq:minimal Hamiltonian}
\end{align}
Here, $\Psi_i=(p_{z,i\uparrow},p_{z,i\downarrow},d_{xy,i\uparrow},d_{xy,i\downarrow})$
collects the $p_z$- and $d_{xy}$-orbital operators.
The Pauli matrices $(\tau_0,\tau_x,\tau_y,\tau_z)$ and
$(s_0,s_x,s_y,s_z)$ act in orbital and spin space, respectively.
$\hat{\bm M}(\psi)$ is the unit vector specifying the magnetization
direction, $\bm{d}_{ij}$ is the
displacement vector from site $i$ to site $j$, and $\langle
ij\rangle$ and $\langle\!\langle
ij\rangle\!\rangle$ denote first-nearest-neighbor and second-nearest-neighbor
pairs with body-centered cubic structure, respectively. We set
$\varepsilon_0=0$,
$\Delta_0=1.0$, $t_0=-1.0$, and
$\Delta=1.0$, and vary $t^{(1)}_{\mathrm{T}}\in[0,0.2]$ and
$t^{(2)}_{\mathrm{T}}\in[0,0.14]$.

\begin{figure}[H]
  \centering
  \begin{subfigure}[b]{0.45\linewidth}
    \centering
    \includegraphics[width=\linewidth]{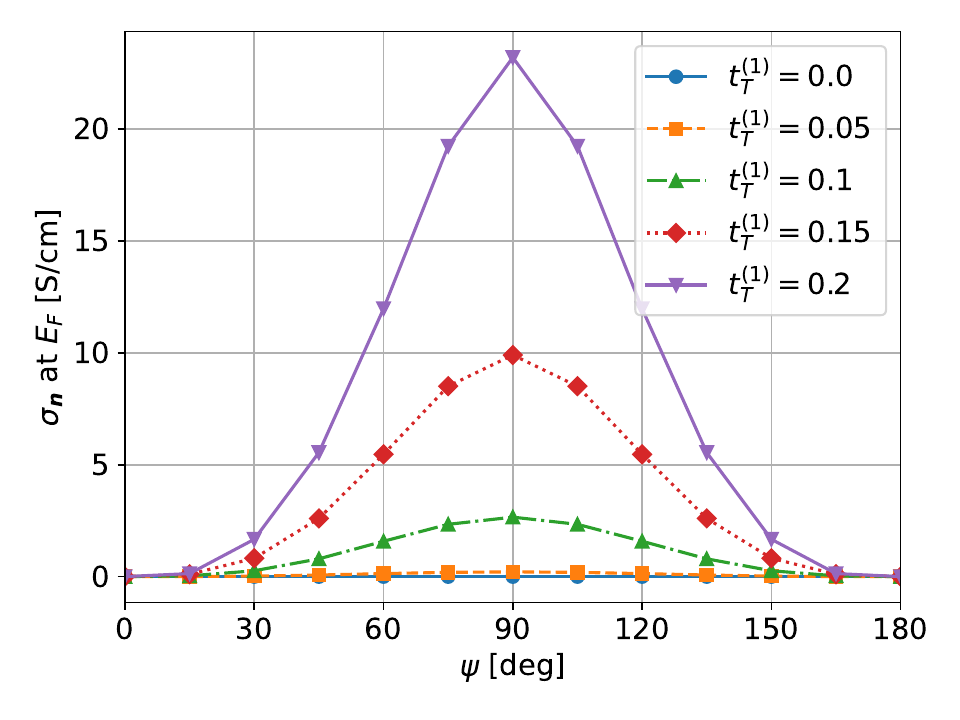}
    \caption{$\sigma_{\bm n}$ in the $(103)$ plane with
    $t^{(2)}_{\mathrm{T}}=0$ as $t^{(1)}_{\mathrm{T}}$ varies.}\label{fig:t1}
  \end{subfigure}\hfill
  \begin{subfigure}[b]{0.45\linewidth}
    \centering
    \includegraphics[width=\linewidth]{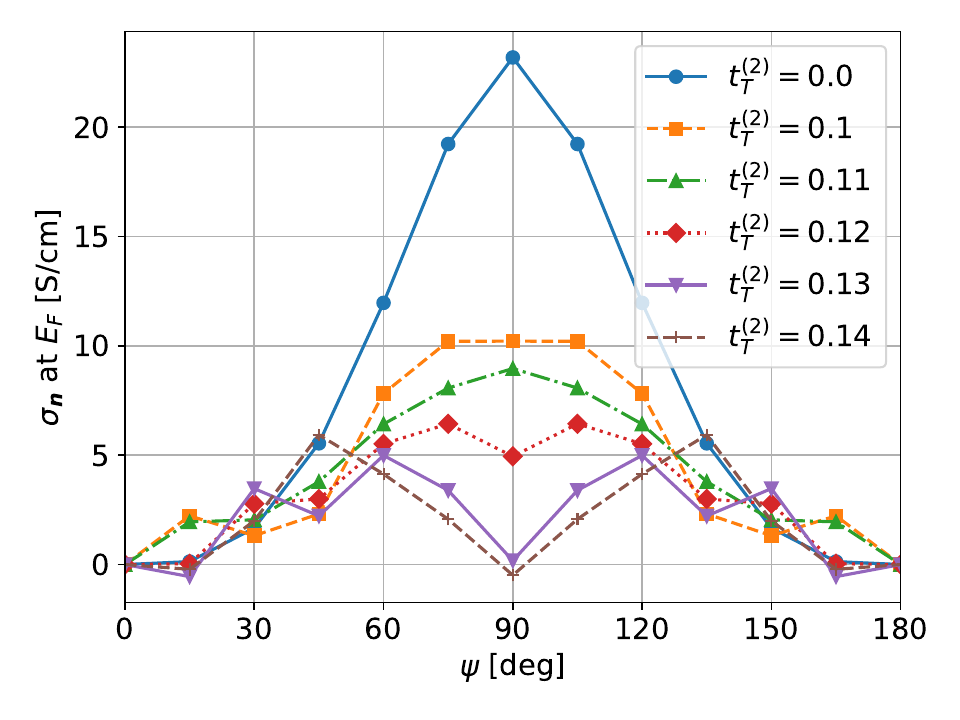}
    \caption{$\sigma_{\bm n}$ in the $(103)$ plane with
    $t^{(1)}_{\mathrm{T}}=0.2$ as $t^{(2)}_{\mathrm{T}}$ varies.}\label{fig:t2}
  \end{subfigure}
  \caption{Out-of-plane AHC $\sigma_{\bm n}$ during magnetization
    rotation in the $(103)$ plane: (a) for $t^{(1)}_{\mathrm{T}}
    \in [0,0.2]$
    with $t^{(2)}_{\mathrm{T}}=0$; (b) for $t^{(2)}_{\mathrm{T}}
    \in [0,0.14]$
  with $t^{(1)}_{\mathrm{T}}=0.2$.}\label{fig:model_103}
\end{figure}

Figure~\ref{fig:model_103} shows the angular dependence of the
out-of-plane AHC component $\sigma_{\bm n}$ during magnetization
rotation in the $(103)$ plane, calculated using the minimal model
Hamiltonian.
In Fig.~\ref{fig:model_103}(\subref{fig:t1}), we vary the first-nearest-neighbor
magnetic-toroidal hopping $t^{(1)}_{\mathrm{T}}$ with $t^{(2)}_{\mathrm{T}}=0$,
whereas in Fig.~\ref{fig:model_103}(\subref{fig:t2}), we vary the
second-nearest-neighbor magnetic-toroidal hopping
$t^{(2)}_{\mathrm{T}}$ with $t^{(1)}_{\mathrm{T}}=0.2$.
From Fig.~\ref{fig:model_103}(\subref{fig:t1}), we find that
$\sigma_{\bm n}$ vanishes at $t^{(1)}_{\mathrm{T}}=0$ and increases in
magnitude with $t^{(1)}_{\mathrm{T}}$, without changing its overall shape.
By contrast, Fig.~\ref{fig:model_103}(\subref{fig:t2}) shows that increasing
$t^{(2)}_{\mathrm{T}}$ deepens the valley of $\sigma_{\bm n}$
around $\psi=90^{\circ}$.
This indicates that
second-nearest-neighbor magnetic-toroidal hopping is the primary
driver of the valley-like profile in $\sigma_{\bm n}$.

\section{Discussion}
This study demonstrates that the TRS-Wannier framework for
magnetization rotation is effective not only for calculating magnetic
anisotropy energy, as shown in the previous study
\cite{Saito2024-vw}, but also for
analyzing AHE anisotropy.
By combining this framework with multipole decomposition using SAMB,
we demonstrate that the symmetry of the resulting Hamiltonians can be
quantitatively evaluated.
Complete SAMB decomposition typically requires millions of basis
matrices, which makes the interpretation of individual multipole
components challenging.
Nevertheless, the decomposition coefficients can serve as
fingerprints unique to each material and magnetic ordering,
with potential applications as descriptors in machine learning.

Notably, significant contributions from higher-order magnetic and
magnetic toroidal multipoles appear not only in systems with $f$-electrons,
such as \ce{UNi4B} \cite{Mentink1994-vc, Oyamada2007-nb,
Saito2018-hc, Ishitobi2023-fj},
\ce{Ce3TiBi5} \cite{Motoyama2018-fo, Shinozaki2020-eu,
Shinozaki2020-yj,Hayami2022-ci}
, but also in the seemingly simple bcc \ce{Fe}.
In particular, we quantitatively demonstrate that high-rank
components—including magnetic octupoles, magnetic toroidal 16-poles,
and magnetic 32-poles—can contribute with strengths comparable to
those of magnetic dipoles.
We further find that the contribution
from magnetic toroidal 16-poles acts with the opposite sign relative
to the other dominant terms.
These findings indicate that the deviations from the angular
formulas, which are derived from an expansion in the magnetization
direction based on the spin group theory, call for the inclusion of
higher-order multipolar terms and/or an explicit treatment of orbital
magnetization arising from magnetic-toroidal hopping processes that
simultaneously change spin and orbital states.

Building on this observation, we show that applying strain can
control the magnetic-toroidal 16-pole contribution and even reverse
the sign of the IAHE. These results establish magnetoelastic
engineering as a viable strategy to manipulate high-order multipoles
in ferromagnets and may offer a promising route toward spintronics applications,
where tunable AHC magnitude and angular dependence can be exploited
for device design.

\section{Conclusion}
We have introduced a microscopic framework that combines time-reversal
symmetric Wannier functions with a symmetry-adapted multipole basis
to analyze the in-plane anomalous Hall effect in ferromagnets. By
decomposing first-principles Hamiltonians into multipole components
and rotating magnetization rank by rank, the method quantifies
symmetry fidelity and identifies the channels that control the
angular dependence of the conductivity. Applied to body-centered
cubic iron, it reproduces the angle dependence from density
functional theory and shows that high-rank magnetic and magnetic
toroidal multipoles are comparable in magnitude to magnetic dipoles
and shape the valley-like response. Uniaxial strain along the [103] direction
lowers symmetry, opens additional magnetic toroidal channels, and
enables tunable reshaping of the angular response, including complete
sign inversion under modest tensile strain. These results
establish multipole-resolved Hamiltonian engineering as a practical
route to predict and control the in-plane anomalous Hall effect and
point to magnetoelastic strategies for tunable spintronic
functionality in simple ferromagnets.

\section{Code availability}
The codes and data will be made available upon request.

\section{Acknowledgement}
This work was supported by
JSPS KAKENHI Grant No.\ 22K03447, and 23H04869,
JST-Mirai Program (JPMJMI20A1),
JST-ASPIRE (JPMJAP2317),
Center for Science and Innovation in Spintronics (CSIS),
Tohoku University and GP-Spin at Tohoku University.

\bibliographystyle{elsarticle-num}
\bibliography{main}
\end{document}